\documentclass{svjour2}                    
\smartqed  
\usepackage{graphicx}
\usepackage[T1]{fontenc}
\usepackage[latin9]{inputenc}
\usepackage{mathrsfs}
\usepackage{amsbsy}
\usepackage{amssymb}
\usepackage{graphicx}
\usepackage{esint}
\PassOptionsToPackage{normalem}{ulem}
\usepackage{ulem}



\begin{document}

\title{A path integral formalism for non-equilibrium Hamiltonian statistical
systems
}


\author{Richard Kleeman       
}


\institute{R. Kleeman \at
              Courant Insitute of Mathematical Sciences \\
              251 Mercer Street, New York, NY 10012 USA\\
              Tel.: +212-998-3233\\
              Fax: +212-995-4121\\
              \email{kleeman@cims.nyu.edu}           
}

\date{Received: date / Accepted: date}

\maketitle

\begin{abstract}
A path integral formalism for non-equilibrium systems is proposed
based on a manifold of quasi-equilibrium densities. A generalized
Boltzmann principle is used to weight manifold paths with the exponential
of minus the information discrepancy of a particular manifold path
with respect to full Liouvillean evolution. The likelihood of a manifold
member at a particular time is termed a consistency distribution and
is analogous to a quantum wavefunction. The Lagrangian here is of
modified generalized Onsager-Machlup form. For large times and long
slow timescales the thermodynamics is of {\"O}ttinger form. The proposed
path integral has connections with those occuring in the quantum theory
of a particle in an external electromagnetic field. It is however
entirely of a Wiener form and so practical to compute. Finally it
is shown that providing certain reasonable conditions are met then
there exists a unique steady-state consistency distribution. 
\end{abstract}

\section{Introduction}
\label{intro}
Around sixty years ago Onsager and Machlup (OM) \cite{onsager1953fluctuations}
proposed a near equilibrium variational principle for determining
the likelihood of time dependent fluctuations in statistical systems
in equilibrium. This principle is formulated as a Wiener path integral
and the associated stochastic process is easily shown to be Ornstein
Uhlenbeck. Formally the path amplitudes $W$ are given by
\begin{eqnarray}
W\left[\lambda(t)\right] & = & C\exp\left[-k^{-1}\int_{0}^{t}\left(\mathbf{\dot{\lambda}}-U\mathbf{\lambda}\right)^{t}g\left(\mathbf{\dot{\lambda}}-U\mathbf{\lambda}\right)dt\right]\label{pathint}\\
 & \equiv & C\exp\left[-k^{-1}\int_{0}^{t}\mathcal{L}\left(\mathbf{\dot{\lambda}},\mathbf{\lambda}\right)dt\right]\\
 & \equiv & C\exp\left[-k^{-1}S\left[\lambda(t)\right]\right]
\end{eqnarray}

where the vector path $\mathbf{\lambda}(t)$ lies in an appropriate
vector space of thermodynamical variables; the $g$ and $U$ are constant
matrices with the former non-negative definite; $k$ is Boltzmann's
constant while $S$ and $\mathcal{L}$ will be referred to as an action
and a Lagrangian respectively. The probability function $p$ with
respect to a thermodynamical variable $\lambda$ at a particular time
$T$ is then given using a path integral over $W$:
\begin{eqnarray*}
p(\lambda_{T}) & = & C\int d\lambda_{0}p(\lambda_{0})K(\lambda_{0},\lambda_{T})\\
K(\lambda_{0},\lambda_{T}) & \equiv & \int_{\begin{array}{c}
\lambda(0)=\lambda_{0}\\
\lambda(T)=\lambda_{T}
\end{array}}W\left[\lambda(t)\right]\mathscr{D}\lambda
\end{eqnarray*}

where the second \uline{path} integral $K(\lambda_{0},\lambda_{T})$
is over \uline{all} paths with endpoints $\lambda_{0}$ and $\lambda_{T}$.
If one fixes $\lambda(0)=\lambda_{0}$ then because $g$ is positive
definite, the action $S$ is minimized by choosing the path which
is a solution of
\begin{equation}
\mathbf{\dot{\lambda}}=U\mathbf{\lambda}\label{OMODE}
\end{equation}
Now for the special case of the action given by (\ref{pathint}) then%
\footnote{With the proviso that $p$ is always Gaussian as was originally assumed
by Onsager and Machlup%
} one can show that this particular path also maximizes $p$ \uline{for
all times} providing $\lambda_{0}$ maximizes $p$ at $t=0$. As noted
by OM it therefore selects the \uline{thermodynamical} path for
the system. 

It is worth observing that this property does not hold for more general
actions and stochastic processes as we shall see in more detail below.
In this contribution, we shall refer to the path which maximizes $p$
for all times as a \uline{thermodynamical} path. The path which
minimizes the action $S$ between any two fixed endpoints and is thus
a solution of the second order Euler-Lagrange equations, we shall
refer to as an \uline{extremal} path. In general it will \uline{not}
be the case that an extremal path between any two points on a thermodynamical
path is in fact a thermodynamical path. This property is however true
for the original OM path integral.

Several relevant questions arise from this seminal formulation
\begin{enumerate}
\item Can this principle be extended to far from equilibrium systems and
if so how exactly? One might hope that a Lagrangian of the same general
form might be possible with the vector $U$ and matrix $g$ being
generalized to being state dependent. This question occupied the attention
of those concerned with general (as opposed to Ornstein Uhlenbeck)
Markov stochastic processes in the 1970s. It was discovered (e.g.
\cite{haken1976generalized} and \cite{graham1977path}) however that
the Lagrangian given in (\ref{pathint}), as well as requiring state
dependent $g$ and $U$, also required the addition of several other
terms. The nature of those terms depended crucially on the time discretization
procedure used to rigorously define the path integral. An attractive
feature of the contribution by \cite{graham1977path} was that the
Lagrangian could be cast into a covariant form with the various quantities
becoming tensors for a Riemannian manifold determined by regarding
$g$ as a metric tensor. Graham and collaborators showed that the
additional terms required in the Lagrangian had a very natural manifold
interpretation and moreover that a natural time discretisation procedure
could be specified to give the derived Lagrangian using renormalization
theory (see \cite{deininghaus1979nonlinear}). 
\item What determines the functional form of the matrices in the Lagrangian
of the original OM path integral and any generalization? In many approaches
to this subject they are simply prescribed empirically. One would
however hope that that they might be derivable from first principles
using the underlying fine-grained dynamics.
\item An original motivation of the OM weight $W$ was as a path generalization
of the Boltzmann principle which relates the probability of a fluctuation
to it's entropy (see \cite{Ein10}). The action $S$ of the path is
therefore argued to be analogous to the entropy of a fixed time fluctuation.
Since the latter can be cast as an information theoretical functional
it would be interesting if a similar functional could be found for
paths. 
\end{enumerate}
The approach to be followed here shall be motivated by an attempt
to answer the above three points. Unsurprisingly the above questions
(particularly the first) have received considerable attention in the
literature where many other approaches aside from those just mentioned
have been proposed. A non-exhaustive list includes \cite{kraichnan1979variational},
\cite{lavenda1979validity}, \cite{eyink1996action}, \cite{taniguchi2007onsager}
and \cite{battezzati2012onsager}. The first and third of these studies
are closest in spirit mathematically to that to be proposed here. 

The approach followed here is based upon the classical approach to
non-equilibrium statistical systems of Zubarev \cite{Zub74} and a
recent extension by Turkington (BT) \cite{turkington2012optimization}.
In that work a set of slow variables $A$ are selected from the system
and non-equilibrium densities of the system averaged over the appropriate
timescale (which we denote by $\Delta t$) are \uline{approximated}
using a maximum entropy principle with constraints provided by the
expectation values of $A$: 
\begin{equation}
\hat{p}(t)=\exp\left[\lambda(t)^{t}A-G(\beta,\lambda)-\beta E\right]\label{trialfamily}
\end{equation}
where $E$ is the system energy%
\footnote{More general invariants than energy of the dynamical system may also
be considered.%
}. Due to their approximating nature $\hat{p}$ are referred to as
trial densities. Their functional form implies that they belong to
the manifold of a general exponential density family (see \cite{ama00}).
Coordinates on such a manifold can be specified using the vector $\lambda$
or the constraining expectation values of the slow variables $\left\langle A\right\rangle $.
Note that the dependence of the trial density on the fast variables
comes solely through the energy function $E$. For many dynamical
systems of interest this implies that the fast and slow variables
are statistically independent. It is reasonably clear physically however
that during equilibration, statistical interaction takes place between
the two sets of variables. This shows the approximate nature of the
trial densities and also implies that initially the equilibration
process from a trial density is a slow one (see BT for a demonstration).
As a consequence trial densities are often also referred to as quasi-stationary.
More discussion on these issues will be given below.

The implicit assumption underlying the present approach then is that
if a sufficiently long time average of the system is taken then the
resulting system density will be close in some sense to particular
members of the trial density family. It is important to stress then
that the central objective of the current approach is to identify
a best \uline{approximating} trial density and from this deduce
good approximating values for $\left\langle A\right\rangle $. The
philosophy adopted is that the actual statistical interaction between
fast and slow variables in non-equilibrium systems is very complex
which implies that only an approximating density may be found. Notice
the contrast in approach to equilibrium studies where the Gibbs density
is commonly assumed to be exact.

The question now arises as to how densities evolve on the particular
slow time scale of interest. For Hamiltonian dynamical systems the
exact densities evolve according to the Liouville equation. Applying
the Liouville propagator to the trial densities results however in
general in a density outside the chosen manifold. One can measure
the discrepancy between this \uline{evolved} density and trial
densities using some appropriate distance functional. Natural choices
for this of course derive from information theory which therefore
allow the discrepancy to be interpreted as an information loss rate.
BT \cite{turkington2012optimization} showed that this loss rate at
a particular time can be formulated as a ``Lagrangian''%
\footnote{Note that this Lagrangian is quite distinct from that applying to
the original Hamiltonian dynamics. It can in some sense be regarded
as a slow variable Lagrangian for the system since $\lambda$ specifies
the slow variable expectation values via a Legendre transform.%
} function $\mathcal{L}_{D}(\dot{\lambda},\lambda)$. The specific
functional form is dependent on the original full Hamiltonian dynamical
system as well as the trial density manifold chosen. This first principles
calculation is discussed in more detail later in this contribution
and in the original BT reference. 

Consider now an experiment in which an initial density is specified
to be \uline{exactly} a trial density. A fixed $\lambda(0)=\lambda_{0}$
is hence assumed. Consider now the set of paths $\lambda(t)$ with
this particular starting point. BT proposed that each such path be
assigned the following ``action'':
\begin{equation}
S\left[\lambda(t)\right]=\int_{0}^{T}\mathcal{L}_{D}(-\dot{\lambda},\lambda)dt\label{BTaction}
\end{equation}

Define now the path minimization function
\begin{equation}
S_{m}(\lambda_{T},T)\equiv\min_{\lambda(T)=\lambda_{T}}S\left[\lambda(t)\right]\label{Brucemin}
\end{equation}

The path achieving such a minimization is, in the terminology introduced
above, an extremal between $\lambda_{0}$ and $\lambda_{T}$ for the
corresponding Lagrangian. The $\lambda_{opt}(t)$ which specify the
coordinates of the best approximating trial density, are now defined
as those values of $\lambda$ which minimize $S_{m}(\lambda,t)$.
It is notable that the path $\lambda_{opt}(t)$ is \uline{not in
general} an extremal path. We comment on this further below as it
is analogous to the difference between a thermodynamical and extremal
path mentioned above in connection with OM theory.

The BT formalism was tested numerically by Kleeman and Turkington
(KT) \cite{kleeman2012nonequilibrium} in a dynamical system which
has often served as a simple model of turbulence: A spectrally truncated
Burgers-Hopf (TBH) model which obeys Hamiltonian dynamics. TBH has
the attractive property that the steady-state statistical density has
been shown numerically to be given by a simple Gaussian Gibbs density%
\footnote{The energy function for TBH is simply the sum of the squares of the
spectral mode amplitudes meaning the Gibbs measure is a Gaussian with
uncorrelated modes and equal variances proportional to the conserved
energy of the system.%
}. The system is also a rather stringent test of the formalism because
the decorrelation timescales of the spectral modes vary inversely
with wavenumber which means that there is not a clean separation between
fast and slow variables. Nevertheless the formalism developed performed
reasonably well in predicting the time evolution of the means of the
slow (low wavenumber) spectral modes both in a situation close to
steady-state and moderately removed from it. In particular after initialisation
with a member of the trial density family, the closure predicted two
qualitative features of the equilibration with high accuracy:
\begin{enumerate}
\item The relaxation time to a steady-state is proportional to the inverse
wavenumber which as noted is proportional to the spectral mode decorrelation
time.
\item The modal relaxation is characterised by an initial ``plateau''
period in which dissipation increases followed by an exponential decay
to a steady-state via an asymptotic dissipation. The plateau period occupied
the same very significant fraction of the relaxation time for all
modes. 
\end{enumerate}
The second property has fundamental implications for the macrostate
description of the system. If the system is restarted at a particular
time after the original start time using the trial density implied
by the the path $\lambda_{opt}(t)$ then, in general, it will follow
a \uline{different} path from that of the original experiment.
Such behaviour occurs in both the direct numerical simulations and
in the theoretical solutions. It occurs theoretically because a period
of increasing dissipation is always evident for a system initialised
with a quasi-stationary density. Consequently the macrostate co-ordinates
$\lambda$ of the system at this evolved later time are insufficient
to \uline{fully} specify the future macrostate evolution. This
reflects the fact that, as was noted above, the identified trial density
is only the best \uline{approximation} to the true density of the
system. What is also the case is that the path of trial densities
most consistent with Liouvilean evolution (the extremal path from
equation (\ref{Brucemin})) is not $\lambda_{opt}(t)$. Again this
is an indication of the inadequacy of choosing just $\lambda_{opt}(t)$
to describe the macrostate at time $t$. It is rather curious that
for the system to equilibrate maximally this approximate behaviour
appears essential. The kind of non-Markovian behaviour just noted
is also an intrinsic part of other non-equilibrium theories such as
that of Mori-Zwanzig (see \cite{zwanzig2001nonequilibrium} and \cite{darve2009computing}). 

In this contribution we shall propose that the macrostate is better
specified using a non-negative \uline{consistency distribution}%
\footnote{Note that we use the terminology \uline{distribution} here to avoid
confusion with the approximating trial densities. The consistency
distribution is a function (or distribution) of the coordinates $\lambda$
which specify the position within the manifold of trial densities.
The densities are defined on the original variables of the Hamiltonian
system.%
} of the trial manifold co-ordinates. When such a distribution is given
at a particular time, the future macrostate evolution of the system
can be computed uniquely. In some respects this approach is analogous
to quantum mechanics where a wave function at a given time is sufficient
via the Schr{\"o}dinger equation to specify the future state of the system.
Indeed the mentioned consistency distribution may be derived in a
natural way from a path integral in the same basic way that a quantum
wave function is derived from a Feynman path integral. The Lagrangian
involved is the $\mathcal{L}_{D}$ discussed above. As usual in statistical
mechanics this path integral is of a Wiener rather than complex Feynman
type. The theory proposed here represents a generalization of the
approach of BT which may be considered as analogous to the classical
limit of the present ``quantum'' theory. The slow time scale $\Delta t$
of the problem plays the analogous role of the quantum $\hbar^{-1}$.
The time varying maximum of the consistency distribution represents
the sequence of trial densities most consistent with Liouville evolution
and the prescribed initial density. We refer to this path as the thermodynamical
path in analogy with OM theory above. In general however unlike OM
theory this path is not an extremal path for $\mathcal{L}_{D}$. 

In the next section we derive the information loss implied in the
choice of a particular time sequence of trial approximating densities.
This loss has an interesting decomposition due to information geometry
into pieces related to reversible and irreversible paths within the
manifold. 

In section 3 we use this derived information loss Lagrangian to propose
a path integral formulation for the problem at hand using a generalized
path Boltzmann principle. This is an idea suggested originally in
a different context by Onsager and Machlup. A very simple pedagogical
example is also given to illustrate fundamental behavior. A physical
interpretation of the consistency distribution is also given. 

In section 4 we compare our path integral with those of OM form using
a Lagrangian transformation due to Roncadelli \cite{roncadelli1992new}.
Mathematically, the present path integral is of a generalized OM form
with the addition to the action of a function at the endpoints of
the path. It is thus similar to the path integral considered by \cite{graham1977path}
and others but there the matrix functions $U$ and $g$ were not determined
from first principles and the terms added to the action were path
dependent. In the limit of large time and large $\Delta t$ the formalism
reduces to the classical OM type and the most consistent or thermodynamical
path becomes one of the type proposed by {\"O}ttinger. 

In section 5 we show that the Lagrangian derived is the same as that
for a non-relativistic particle moving in an external magnetic field
as well as an external potential. The particle moves in a manifold
specified by a metric tensor given by the Fisher information matrix
$g$ of the exponential family assumed. 

In section 6 we consider the Schr{\"o}dinger equation associated with
the proposed path integral. In section 7 we note the similarity and
differences to the Wick rotated electromagnetic path integral of equilibrium
quantum statistical mechanics. In section 8 we consider the associated
(time) transfer operator and show using compact operator theory that
there exists a unique consistency distribution associated with a steady-state.
Section 9 contains a discussion.

\section{Path Liouville discrepancy }
\label{sec:2}

Since we intend invoking a generalized Boltzmann principle in the
next section, we derive here an information theoretic based measure
of the discrepancy of a time sequence of trial densities from Liouvillean
evolution. For more detail the reader is also referred to the earlier
work BT where this idea was first introduced using a somewhat different
approach. 

Suppose we are dealing with a Hamiltonian dynamical system with the
symplectic evolution equation for a general variable given by:
\[
\frac{dF}{dt}=\left\{ F,H\right\} +\frac{\partial F}{\partial t}
\]
where $H$ is the system Hamiltonian and the Poisson bracket is given
by
\begin{equation}
\left\{ A,B\right\} =\left(\nabla A\right)^{t}J\nabla B\label{Poisson}
\end{equation}
with the gradient taken with respect to the dynamical variables and
the matrix $J$ is antisymmetric which ensures the bracket is antisymmetric
with respect to its two arguments. A (smooth) probability density
$p$ on this dynamical system satisfies the Liouville equation
\begin{eqnarray}
\frac{\partial p}{\partial t}+Lp & = & 0\label{Liouvilleequation}\\
Lg & \equiv & \left\{ g,H\right\} \nonumber 
\end{eqnarray}

with the operator $L$ anti-Hermitian with respect to the usual Hilbert
space inner product. 

Consider the anti-Hermitian differential operators

\[
L\equiv-\frac{\partial H}{\partial x_{i}}J_{ij}\frac{\partial}{\partial x_{j}}\qquad T\equiv\frac{\partial}{\partial t}
\]

where $x_{i}$ are the basic (fine grained) dynamical system variables.
We assume that these operators commute i.e. that the gradient of $H$
and $J$ do not depend explicitly on $t$. Denote now a trial density
by $\hat{p}$ and consider various temporal evolutions over a short
interval $\Delta t$ which is however assumed sufficiently long that
unresolved degrees of freedom decorrelate. The evolution according
to the Liouville equation (\ref{Liouvilleequation}) will be
\[
\overline{p}(t+\Delta t)\equiv e^{-\Delta tL}\hat{p}(t)
\]

Now in general%
\footnote{If the trial distribution gives an invariant measure for the system
this will not be the case.%
} this evolved density will lie outside the manifold described by trial
densities. The evolved \uline{trial} density must therefore be
the different density
\[
\hat{p}(t+\Delta t)=e^{\Delta tT}\hat{p}(t)
\]

The information lost $IL$ in assuming $\hat{p}(t+\Delta t)$ when
in fact the density is $\overline{p}(t+\Delta t)$ is simply the relative
entropy $D(*||*)$ of the second density with respect to the first.
We have now the following
\begin{eqnarray}
IL & = & D\left(e^{-\Delta tL}\hat{p}||e^{\Delta tT}\hat{p}\right)\nonumber \\
 & = & \int e^{-\Delta tL}\hat{p}\left(e^{-\Delta tL}\hat{l}-e^{\Delta tT}\hat{l}\right)\nonumber \\
 & = & \left\langle e^{\Delta tL}\left(e^{-\Delta tL}-e^{\Delta tT}\right)\hat{l}\right\rangle _{\hat{p}}\nonumber \\
 & = & \left\langle \left(I-e^{\Delta tL}e^{\Delta tT}\right)\hat{l}\right\rangle _{\hat{p}}\label{il1}\\
 & = & \left\langle \left(I-e^{\Delta t\left(T+L\right)}\right)\hat{l}\right\rangle \label{illast}
\end{eqnarray}

with $\hat{l}\equiv\log\hat{p}$. On the second line we are using
the fact that an arbitrary function of $p$ also obeys the Liouville
equation (\ref{Liouvilleequation}); on the third line we are using
the anti-Hermitean property for $L$; and on the last line we are
using $\left[L,T\right]=0$ and the expectation refers to the trial
density at the start of the propagation interval. Define now the following
useful random variable $R$ which we call the Liouville residual
\begin{equation}
R(p)\equiv\left(T+L\right)\log p\label{residual}
\end{equation}

Note that for a probability evolving according to the Liouville equation,
$R$ vanishes but will not in general for a $\hat{p}$ constrained
to lie within the trial density manifold. A general random variable
$F$ can be shown (see Appendix) to satisfy the following evolution
equation
\[
\frac{\partial\left\langle F\right\rangle }{\partial t}-\left\langle LF\right\rangle =\left\langle TF+FR\right\rangle 
\]

from which we deduce (setting $F=1$) firstly that
\begin{equation}
\left\langle R\right\rangle =0\label{Rfirstmoment}
\end{equation}

and secondly (setting $F=R$) that
\begin{equation}
\left\langle \left(T+L\right)R\right\rangle =-\left\langle R^{2}\right\rangle \label{Rsecondmoment}
\end{equation}

Returning now to equation (\ref{illast}) we expand the exponential
operator as a Taylor series. The terms in $\Delta t$ of order zero
and one vanish due to cancellation and equation (\ref{Rfirstmoment})
while the order two term remains and using (\ref{Rsecondmoment})
we derive the remarkably simply second order approximation 
\[
IL=\frac{\left(\Delta t\right)^{2}}{2}\left\langle R^{2}\right\rangle +O(\left(\Delta t\right)^{3})
\]
Thus the information loss to lowest order is simply proportional to
the variance of the Liouville residual $R$. It is worth observing
that this loss is quadratic in the time interval $\Delta t$ which
is consistent with the relative entropy geometrically being a distance
squared (see \cite{ama00}). 

In order to make further progress beyond this general equation we
now specify the trial density manifold $\mathcal{T}$. We identify
a subset of functions $A$ (assumed a vector) from the dynamical system
which we label as the \emph{resolved }(or coarse grained) variables.
In general these will be functions of the slow variables for the dynamical
system. Secondly we assume that steady-state densities are of a Gibbs
type and for simplicity we assume that the only invariant involved
here is the energy. The general trial density is then deduced by minimizing
the relative entropy with respect to the Gibbs density under the assumption
that the resolved variable expectations are known. They therefore
take the form as discussed in the previous section
\begin{equation}
\hat{p}(t)=\exp\left[\lambda(t)^{t}A-G(\beta,\lambda)-\beta E\right]\label{exponential}
\end{equation}

where $E$ is the energy of the system which we are assuming is one
of the resolved variables and satisfies $LE=0$. Note also that $G$
normalizes the distribution and the partition function $Z=\exp G$.
In addition there is a one to one relationship between the co-ordinates
of the manifold $\lambda$ and the expectation values $a$ of the
chosen $A$. Either can serve as co-ordinates for the trial distribution
manifold and are related by a Legendre transform (see, for example,
\cite{ama00}). With this specification it is easy to calculate $R$
as 
\[
R=\dot{\lambda}^{t}(A-a)+\lambda^{t}LA
\]

where the overdot denotes a time derivative and hence that 
\begin{eqnarray}
IL & = & \frac{\left(\Delta t\right)^{2}}{2}\left(\dot{\lambda}^{t}g\dot{\lambda}-2\dot{\lambda}^{t}\left\langle LA\right\rangle +\phi\right)+O(\left(\Delta t\right)^{3})\label{ILnew}\\
\phi & \equiv & \lambda_{i}\left\langle LA_{i}LA_{j}\right\rangle \lambda_{j}\nonumber \\
g_{ij} & \equiv & \left\langle \left(A_{i}-a_{i}\right)\left(A_{j}-a_{j}\right)\right\rangle \nonumber 
\end{eqnarray}

The matrix/tensor $g$ here is the Fisher information matrix which
plays a central role as a Riemannian metric tensor in the field of
information geometry (see \cite{ama00}). We have also used the following
identity derived in Appendix A:
\[
\left\langle LA_{i}\right\rangle =-\lambda_{j}\left\langle \left(A_{i}-a_{i}\right)LA_{j}\right\rangle 
\]

There is an interesting decomposition of the information loss $IL$
which relates both to reversible thermodynamics and to the basic information
geometry we are considering. The entropy $S$ along a general trajectory
may easily be computed as
\begin{eqnarray*}
S & = & -\left\langle \log\hat{p}\right\rangle =-\lambda^{t}a+G+\beta u\\
u & \equiv & \left\langle E\right\rangle 
\end{eqnarray*}
Taking the time derivative we obtain (see Appendix A)
\begin{equation}
\dot{S}=-\lambda^{t}g\dot{\lambda}+\beta\dot{u}\label{entropyrate}
\end{equation}
 Suppose we now define a particular trajectory in our trial distribution
manifold which satisfies the following first order differential equation:
\begin{equation}
\frac{d\tilde{\lambda}}{dt}=g^{-1}\left\langle LA\right\rangle \label{reversible}
\end{equation}
where $\tilde{\lambda}$ is used to distinguish this particular trajectory
from a general trajectory which we write simply as $\lambda$. Obviously
a specification of co-ordinates for a given time will then specify
the particular trajectory given equation (\ref{reversible}). Combining
equations (\ref{entropyrate}) and (\ref{reversible}) we obtain for
this particular trajectory that (see Appendix A):
\[
\dot{S}=\beta\dot{u}
\]

which is the usual expression for reversible entropy change in an
open system with varying mean energy. We therefore identify the particular
trajectory above as a \uline{reversible} trajectory. The information
loss along this reversible trajectory can be computed simply by substituting
(\ref{reversible}) into (\ref{ILnew}) giving to second order accuracy
\begin{equation}
IL_{rev}=\frac{\left(\Delta t\right)^{2}}{2}\left(\phi-\left\langle LA\right\rangle ^{t}g^{-1}\left\langle LA\right\rangle \right)\label{ILrev}
\end{equation}

Finally we can compute $IL_{irr}$ the relative entropy between a
reversible and a general irreversible trajectory within our manifold.
Since both lie within the manifold their relative entropy can be calculated
to second order accuracy by the following well known relation in information
geometry between relative entropy and the Fisher metric (see \cite{ama00}):
\[
D(\hat{p}(\lambda)||\hat{p}(\lambda+\epsilon v))=\frac{\epsilon^{2}}{2}v^{t}gv+O(\epsilon^{3})
\]
thus to second order accuracy we obtain, using the defining relation
for a reversible trajectory 
\begin{equation}
IL_{irr}=\frac{\left(\Delta t\right)^{2}}{2}\left(\dot{\lambda}-g^{-1}\left\langle LA\right\rangle \right)^{t}g\left(\dot{\lambda}-g^{-1}\left\langle LA\right\rangle \right)\label{irreversible}
\end{equation}

It is now trivial to verify the following interesting relation between
various information losses which is accurate to second order:
\begin{equation}
IL=IL_{rev}+IL_{irr}\label{decomp}
\end{equation}

The non-negativity of relative entropy now shows that over the timestep
$\Delta t$ the information loss to second order can be minimized
to $IL_{rev}$ by choosing the reversible trajectory. The endpoint
of the reversible trajectory can thus be viewed as a projection%
\footnote{Strictly this identification as a projection is precise only in the
limit as $\Delta t\rightarrow0$%
} from the fully Liouvillian evolved initial trial distribution back
into the trial manifold. $IL_{irr}$ represents the information loss
in not choosing this infinitesimally optimal reversible trajectory
while $IL_{rev}$ represents the minimum possible information loss
for all trajectories. The full situation is depicted schematically
in Figure 1.
\begin{figure}

\includegraphics[width=0.75\textwidth]{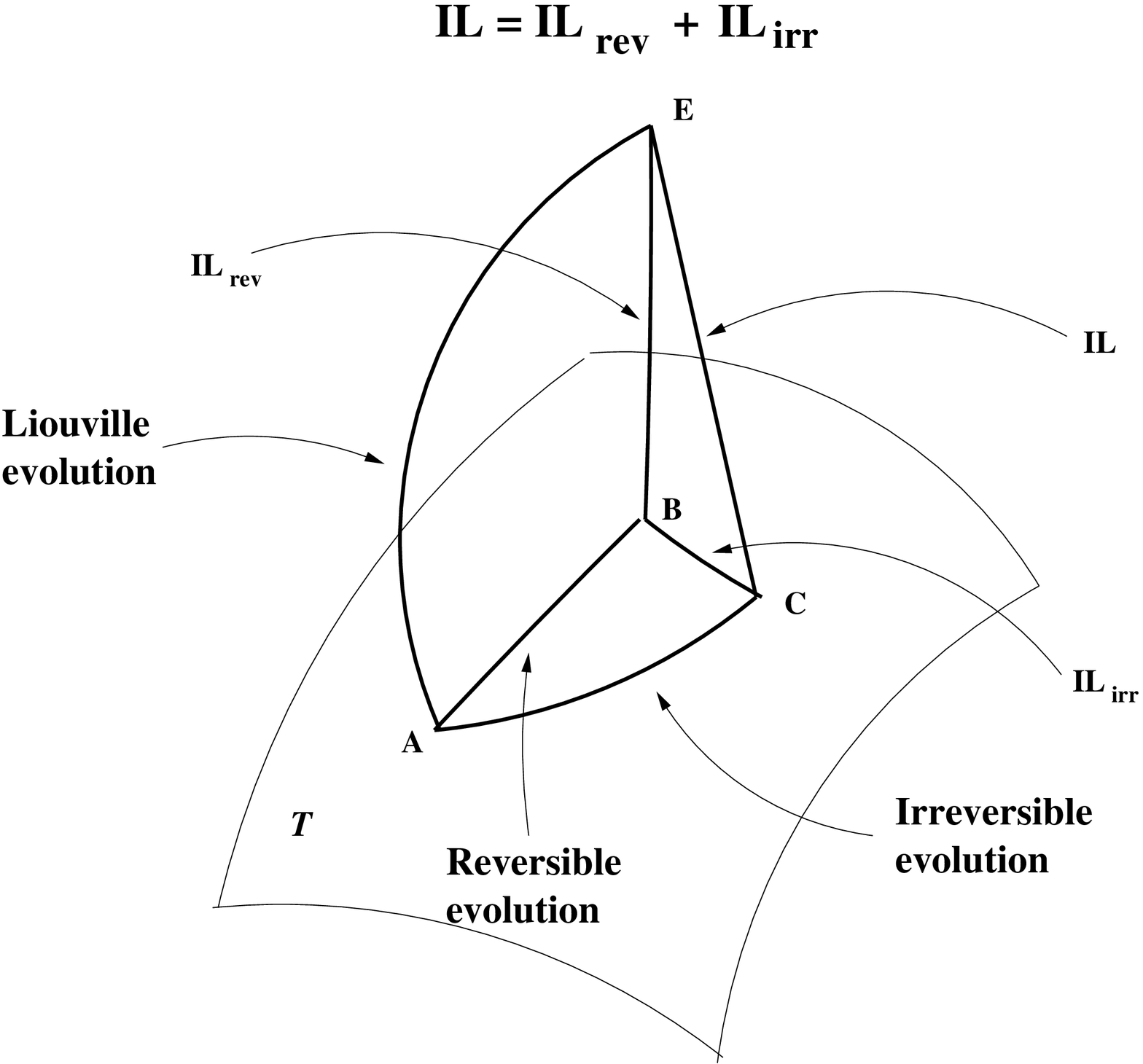}\caption{Information loss decomposition to second order accuracy. Liouville
evolution takes a distribution $A$ in the trial manifold $\mathcal{T}$
to the distribution $E$ which lies outside $\mathcal{T}$. The ``nearest''
distribution in $\mathcal{T}$ to $E$ is $B$ in the sense that it
has minimum relative entropy $D(E||*)$. Thus $B$ may be considered
a projection of $E$ into $\mathcal{T}$. A general distribution $C$
in the trial manifold differs from $B$ by the relative entropy of
$D(B||C)$ and the projective nature of $B$ ensures that $D(E||C)=D(E||B)+D(B||C)$
or $IL=IL_{rev}+IL_{irr}$. This relation is known in information
geometry as a Pythagorean relation since relative entropy for small
displacements within the manifold $\mathcal{T}$ can be regarded as
a squared distance.}
\label{fig:1}       
\end{figure}
It should be clear however that if one chooses a large number of timesteps
the reversible trajectory will no longer in general minimize information
loss since $IL_{rev}$ clearly depends on the trajectory chosen and
there are usually irreversible trajectories which result in smaller
values of this quantity at a given time than that occuring on the
reversible trajectory. The relation (\ref{decomp}) has been discussed
at length in information theoretic contexts (see \cite{ama00} and
\cite{cov91} Chapter 11) where it is referred to as the relative
entropy Pythagorean relation since this functional is best viewed
as a distance squared. Note that the decomposition above was first
discussed in BT in a somewhat different context. Here we have emphasized
the information theoretic perspective for reasons that will become
apparent when we turn to the path integral formalism in the next section. 

The relevant dynamical object of interest is, of course, a long time
path in the trial distribution manifold. The total informational discrepancy
of interest is then simply proportional to the sum of each $IL$ along
the time interval. Mathematically it is convenient to pass partially
to the infinitesimal time limit in which case this becomes the time
integral of a Lagrangian i.e. the action
\begin{eqnarray}
S & = & \Delta t\int_{0}^{T}\mathcal{L}dt\label{limiting-1}\\
\mathcal{L} & \equiv & \frac{1}{2}\left(\dot{\lambda}^{t}g\dot{\lambda}-2\dot{\lambda}^{t}M+\phi\right)\label{TurlLagrangian}\\
M & \equiv & \left\langle LA\right\rangle 
\end{eqnarray}

Notice that the timestep $\Delta t$ enters into the final result
as a consequence of the information loss (relative entropy) being
geometrically a distance squared. 

Finally it is worth observing that a somewhat more general formulation
than above has been proposed and tested in BT and KT. There the two
parts of the information loss $IL_{rev}$ and $IL_{irr}$ are weighted
differently. This was in recognition of the fact that the formalism
being considered is an idealisation in two important respects:

Firstly in reality the fast and slow time scales are never cleanly
separated. Secondly there is arbitrariness in how resolved variables
$A$ are selected from functions of the system slow variables. In
the two concrete dynamical systems examined to date in KT and BT it
has been found convenient to choose the weighting somewhat differently
than the unit ratio in (\ref{decomp}). For the truncated Burgers-Hopf
turbulence system investigated in KT the optimal weighting for agreement
with direct numerical simulations of the full system was found by
increasing the weight of $IL_{rev}$ to around $1.3$. In that case
however the set of resolved variables was simply the slow, small wave
number spectral modes. Since in direct simulations of the full system,
slow mode \uline{variance} variation is apparent, such a set of
resolved variables may well be too restrictive and the set should
be extended to include quadratic functions of the slow modes. 

Notice that if we ignore $IL_{rev}$ altogether in the decomposition
then it is easily seen that the reversible trajectory results from
minimization. These issues will be examined in more depth in future
publications by considering the convergence issue of larger sets of
resolved variables $A$ and also by analyzing a range of different
dynamical systems.

\section{Path integral formulation}
\label{sec:3}

In the previous section we have associated an arbitrary differentiable
path in the manifold $\mathcal{T}$ of trial densities with a non-negative
information loss. Thus from this calculation there exists an obvious
way of weighting paths which is entirely analogous to the OM case
discussed earlier. We are however not interested in path optimality
directly. Instead we are interested in best describing the statistical
system \uline{at a particular time} and hence identifying a thermodynamical
path for the system. 

On the time interval $\left[0,T\right]$ consider the set $\Lambda$
of (differentiable) paths $\lambda(t)$ with fixed endpoints $\lambda(0)=\lambda_{0}$
and $\lambda(T)=\lambda_{T}$. It seems reasonable that the consistency
attached to $\lambda_{T}$ should be some function of the information
loss of \uline{all} the members of $\Lambda$. How should such
a function be constructed however? Clearly paths with small information
loss should contribute more than those with a larger loss since they
are more consistent with Liouvillean evolution. Evidently there are
many possible ways in which this could be achieved however a very
natural way is provided by a Wiener path integral in the manner of
Onsager and Machlup. They argued that their action should play the
role among paths that entropy does for fluctuations. The action we
have defined in the previous section is a path information loss which
is analogous to entropy. We adopt therefore a path Boltzmann principle
and assign a non-negative Wiener path measure by 
\begin{equation}
W\left[\lambda(t)\right]=C\exp\left[-\Delta t\int_{0}^{t}\mathcal{L}(\dot{\lambda},\lambda)\right]\label{Wienermeasure}
\end{equation}

Note that during any time step $\Delta t$ there is an information
loss $IL$ for assuming any step within the trial manifold rather
than Liouville evolution. This is converted to a consistency weight
using a Boltzmann principle. These weights are then multiplied up
along a chosen manifold path to form the non-negative measure $W$. 

The consistency distribution $\psi$ for $\lambda_{T}$ is now simply
the ``sum'' of path measures for all members of $\Lambda$ i.e.
it is simply the path integral:
\[
\psi(\lambda_{T})=K(\lambda_{0},\lambda_{t})\equiv\int_{\begin{array}{c}
\lambda(0)=\lambda_{0}\\
\lambda(t)=\lambda_{t}
\end{array}}W\left[\lambda(t)\right]\mathscr{D}\lambda
\]

Clearly as $\Delta t\rightarrow\infty$ only the extremal path from
$\Lambda$ contributes to the path integral since the relative
weight of all other paths becomes small. Thus in this limit our consistency
distribution is simply $C\exp(-\Delta tS_{m})$ where $S_{m}$ is
the extremal action and the optimal choice for $\lambda_{T}$ is provided
by the value minimizing $S_{m}$ and we return to the formalism proposed
in BT. In general however the slow timescale $\Delta t$ will be finite
and of physical significance to the problem being considered. This
means that the consistency distribution will be a function of all
paths leading to $\lambda_{T}$ not simply the extremal. The difference
between the BT formalism and the present generalization is entirely
analogous to the difference between classical and quantum mechanics.

The consistency distribution at time $t_{2}$ may then be defined,
as in most path integral approaches, as the integral of this amplitude
muliplied by the consistency distribution at $t_{1}$. There remains
then the issue of identifying the appropriate consistency distribution
at the initial time. Now obviously we can, as a practical matter,
specify the initial probability density exactly from the manifold
of trial distributions. Given this knowledge the obvious choice for
an initial consistency distribution is simply a Dirac delta function
centered on the manifold point chosen. One may evidently consider
other choices for the initial density which do not lie within the
trial distribution manifold. We defer consideration of that case to
a later publication.

\subsection{A simple pedagogical example with macrostate ambiguity and plateau
behaviour.}

In order to gain some concrete insight into the formalism proposed
above we now consider the simplest relevant case namely that for exponentially
damped relaxation to a steady-state. Analysis in KT indicates that a
straightforward generalization of this system is relevant to the near
to steady-state relaxation of the TBH system. As we shall see below this
very simple system exhibits the macrostate ambiguity and plateau behaviour
discussed in the introduction. The Lagrangian here is given by 
\[
2\mathcal{L}=\dot{u}^{2}+\kappa^{2}u^{2}
\]

which has the Euler Lagrange equation
\[
\ddot{u}=\kappa^{2}u
\]

The solution of these equations with fixed endpoints is 
\begin{eqnarray}
u(t) & = & Ae^{\kappa t}+Be^{-\kappa t}\nonumber \\
B=\frac{1}{2}\frac{u(0)e^{\kappa T}-u(T)}{\sinh(\kappa T)} &  & A=u(0)-B\label{ELeq}
\end{eqnarray}

Note the importance of not just the damped solution but also the exponential
growing one. The action with respect to this extremal can now be computed
with a little algebra
\begin{eqnarray*}
S_{e}(\kappa,T) & = & \int_{0}^{T}\mathscr{\mathcal{L}}(\dot{u},u)dt\\
 & = & \frac{\kappa}{2}\left[\coth(\kappa T)\left(u(0)^{2}+u(T)^{2}\right)-2u(0)u(T)\mathrm{csch}(\kappa T)\right]
\end{eqnarray*}

which is a very standard result in path integral theory (see e.g.
\cite{feynman1965quantum} equation (10.44)). Suppose we fix $u(0)$
then this action is minimized by a $u_{m}(T)$ satisfying
\begin{equation}
u_{m}(T)=u(0)\mathrm{sech}(\kappa T)\label{maxbruce}
\end{equation}

which satisfies the first order differential equation 
\[
\dot{u}_{m}=-k\tanh\left(\kappa t\right)u_{m}
\]

In otherwords the linear dissipation coefficient increases from zero
to $\kappa$ as time proceeds. 

If we set $\kappa =i\omega$ the system above becomes a standard harmonic
oscillator for which the Feynman path integral is well known \cite{feynman1965quantum}
to be simply 
\[
K_{F}(u(0),u(T))=C\exp\left(\frac{i}{\hbar}S_{e}(i\omega,T)\right)
\]

which implies that the Wiener path integral for this problem is 
\begin{equation}
K\left(u(0),u(T)\right)=C\exp\left(-\Delta tS_{e}(\kappa,T)\right)\label{pathharmonic}
\end{equation}

which is a Gaussian density whose peak is obviously given by equation
(\ref{maxbruce}). Thus in this very simple case the thermodynamical
path does not depend on the slow timescale $\Delta t$ since it is
obtained by simply minimizing the extremal action between the fixed
starting point and all endpoints. For higher order realistic Lagrangians
appropriate for significantly non-equilibrium situations however it
is very important to emphasize that a simple equation of the form
(\ref{pathharmonic}) will not hold. The thermodynamical path then
will indeed depend on $\Delta t$ and it will not be possible to obtain
it by minimizing the action for extremal paths. 

Suppose now we set $u(T)=u_{m}(T)$ then it is easy to see from (\ref{ELeq})
and (\ref{maxbruce}) that for $t<T$ we have $u(t)\neq u_{m}(t)$.
. Furthermore if one restarts the system at $u_{m}(t)$ then the future
thermodynamical trajectory differs markedly from the original. This
is illustrated in top panel of Figure 2 for $\kappa =u(0)=1$. Note in both
cases the initial plateau in the equilibration before exponential
decay occurs. This behaviour is qualitatively the same as seen in
DNS simulations of the truncated Burgers turbulence system analyzed
in \cite{kleeman2012nonequilibrium}. This situation suggests intuitively
that the trial density at time $t>0$ can only be an approximation
to the actual density for that time. This can be seen concretely by
computing the consistency distribution which is proportional to $\exp(-\Delta tS_{e})$.
The results are shown in the bottom panel of Figure 2 for $\Delta t=1$
where it is clear that at the restart time there is a rather broad
distribution. 
\begin{figure}
\begin{centering}
\includegraphics[width=0.75\textwidth]{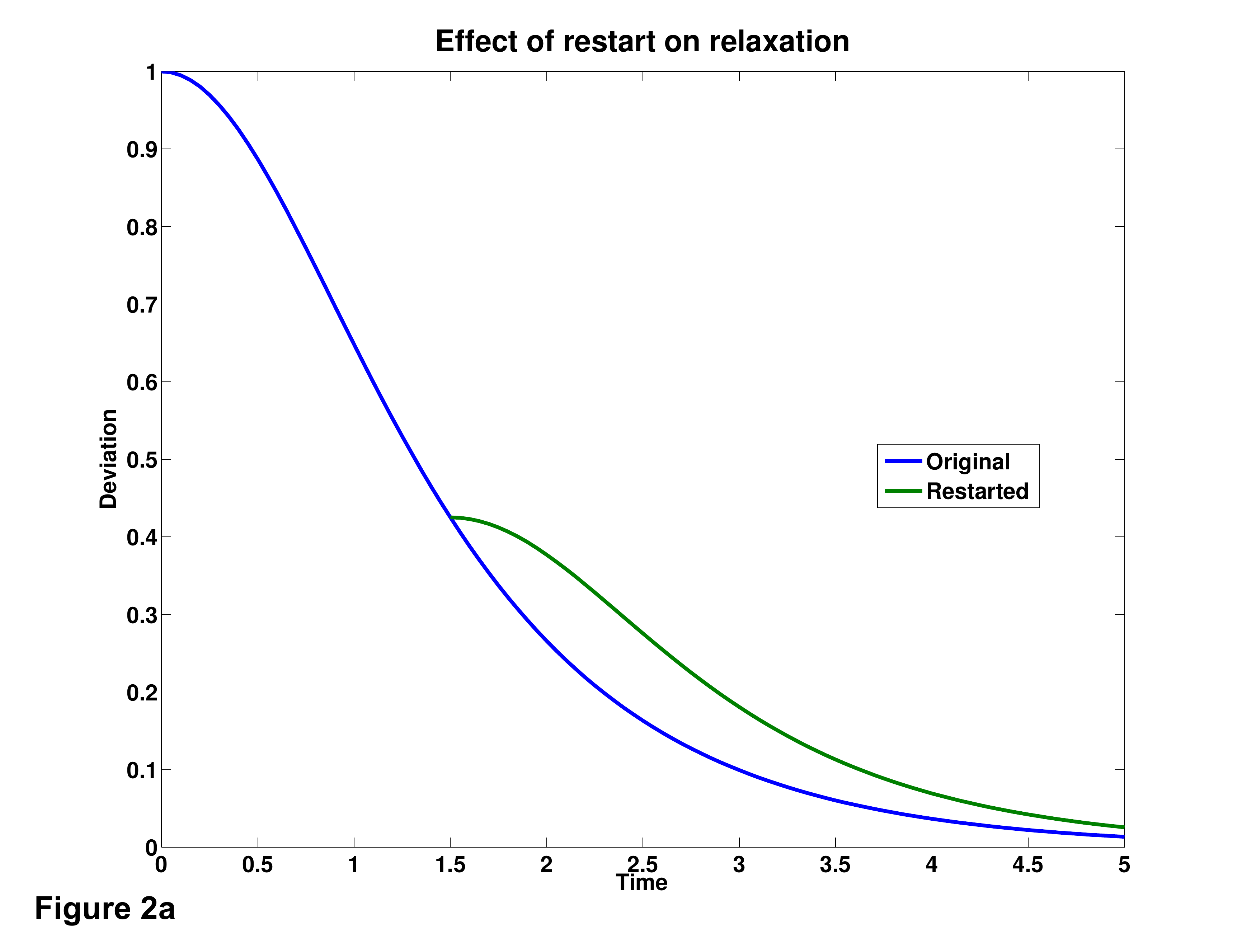}
\par\end{centering}

\begin{centering}
\includegraphics[width=0.75\textwidth]{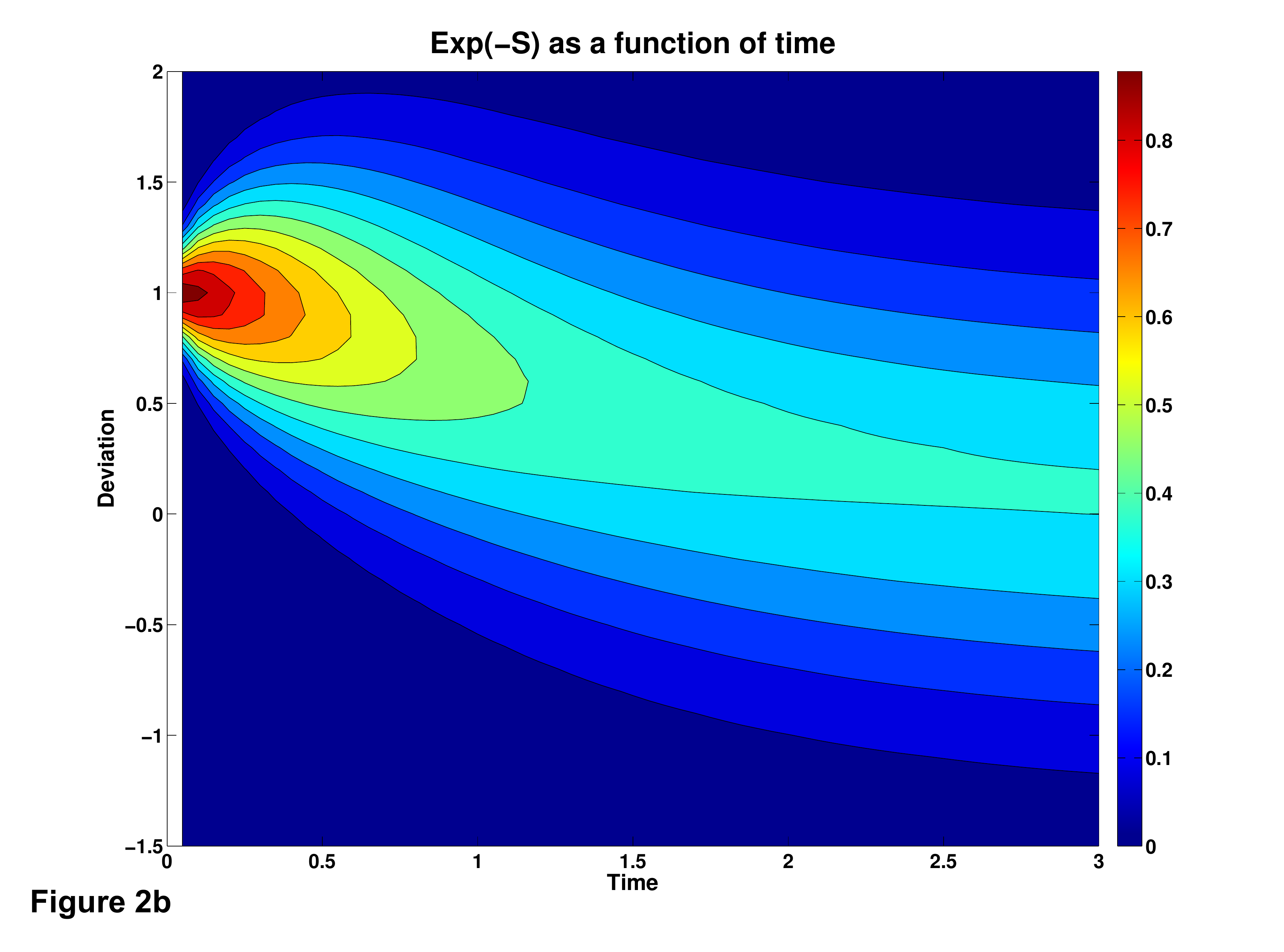}
\par\end{centering}

\caption{Top panel (a): The thermodynamical paths $u_{m}$ are shown for the
cases where the start is at either $t=0$ or at $t= 1.5$. In both
cases the start values are on the thermodynamical path started at
$t=0$ (see text). Note the initial plateau periods in both cases
before exponential decay to a steady-state occurs. Bottom panel (b):
The weights at various times for the case $\Delta t=1$. Note the (Gaussian)
spread at $t=1.5$ where the restart occurs.}

\label{fig:2}       
\end{figure}
Another interesting aspect of the solutions is that the thermodynamical
path $u_{m}(t)$ does not correspond with the extremal path $u(t)$
between any two points on the thermodynamical path (see section 1).
This is demonstrated in Figure 3 for an endpoint close to the steady-state
relative to the initial conditions. Thus the path with the minimal
total information loss does not correspond with the sequence $u_{m}(t)$.
The latter appears ``more realistic'' in that it exhibits the spinup
character universally noted in DNS solutions and discussed in Section
1. 
\begin{figure}
\includegraphics[width=0.75\textwidth]{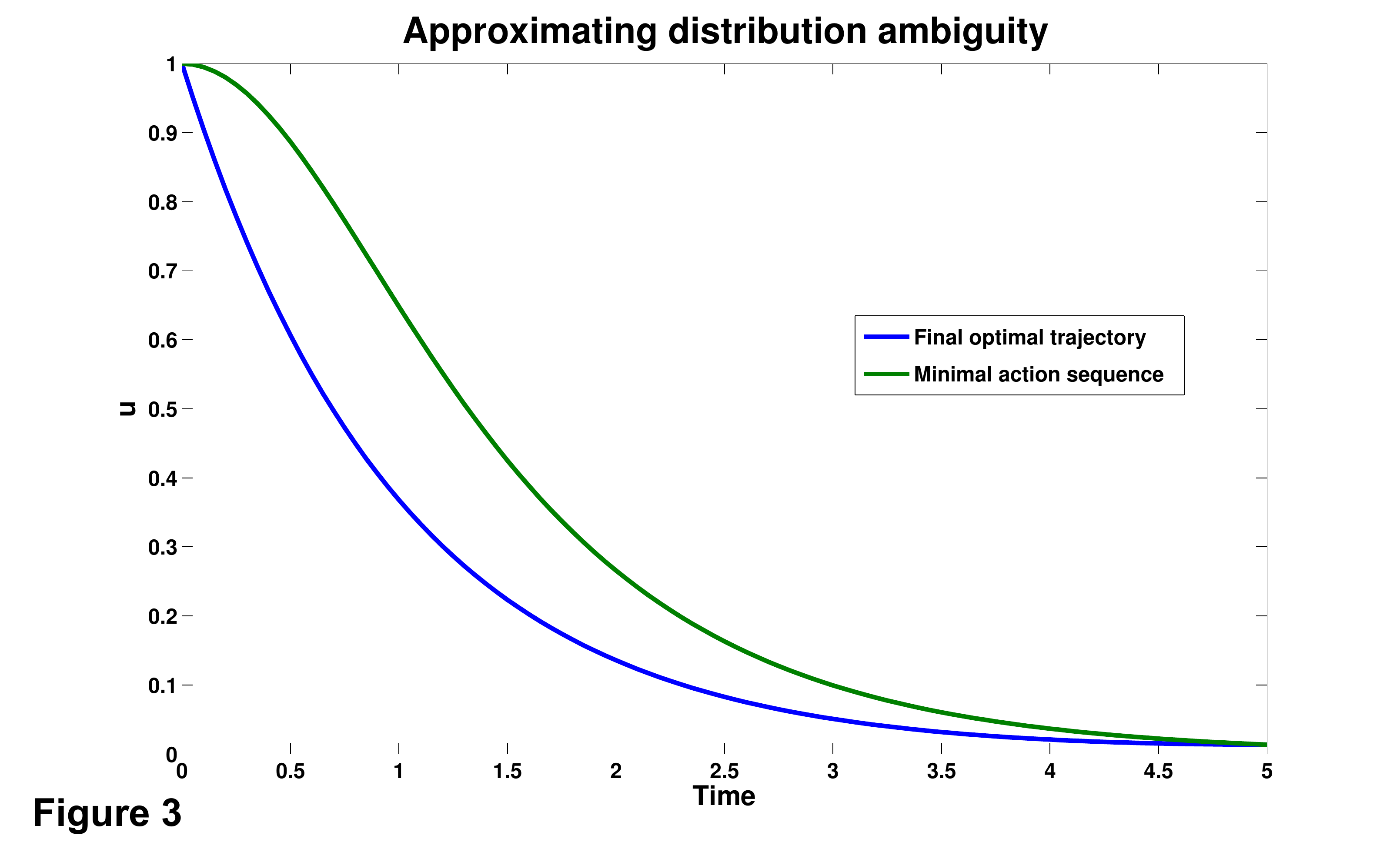}

\caption{Displayed are the thermodynamical path together with the extremal
path between the thermodynamical start and end points (for $t=5$).
Recall that this latter path gives the minimum information loss between
these endpoints.}
\label{fig:3}       
\end{figure}

\subsection{The physical interpretation of the consistency distribution}

The non-negative distribution $\psi(\lambda(t))$ represents how consistent
a trial density is with our knowledge of the initial density and the
fact that the true density obeys the Liouville equation. It depends
importantly therefore on our assumption of which trial density family
is appropriate for the problem. 

In some respects this situation is analogous to the likelihood function
of mathematical statistics. There a parametric statistical family
is selected based on assumptions concerning the nature of the problem
at hand. A sample is then obtained and the likelihood of a particular
choice of parameters deduced. The maximum likelihood set of parameters
then represents the best available choice from the statistical family
given the sample data available. Of course a different sample results
in a different likelihood function which is quite different to the
situation here where the consistency distribution is fixed once the
manifold and initial density are specified. In the statistical modeling
case it is also possible to deduce from the nature of the likelihood
function what the uncertainty of the model parameters are. This is
achieved using the Fisher information matrix. In that case however
it is required to assume that the unknown true population is actually
drawn from a particular family density. In the present situation we
know that the true density very likely does not belong exactly to
the trial density family since interaction takes place statistically
between fast and slow variables during equilibration. Such interaction
is typically complex and not able to be modelled exactly using a trial
density.

Another situation analogous to the present one is provided by the
quantum wavefunction. There only the complex modulus is of direct
experimental significance. The complex phase information is however
of relevance in describing the dynamical evolution of the physical
state via a Schr{\"o}dinger equation of some kind. This situation is exactly
analogous to the present theory where the consistency distribution
\uline{maximum} is directly relevant for defining the thermodynamical
path for the system but the rest of the distribution again is relevant
for describing the dynamical evolution of the macrostate. In future
work the author will explore whether more than simply the maximum
of the consistency distribution can be used in defining slow variable
expectation values and their uncertainty.

Another interesting situation occurs when we know the true density
as $t\rightarrow\infty$ to a high degree of accuracy. Often it can
be assumed to be very close to a Gibbs density of some kind. If such
a density is included in the trial manifold then an asymptotic constraint
as well as the initial constraint may be imposed in defining the consistency
distribution.

\section{Transformation to an Onsager Machlup like path integral}
\label{sec:4}

The Lagrangian specified by equation (\ref{TurlLagrangian}) is not
of the same generic OM form of (\ref{pathint}). There exists however
an interesting transformation that illuminates the relationship between
the two which was originally suggested in the quantum context by Roncadelli
\cite{roncadelli1992new}. The transformation can also be viewed as
a gauge transformation in the sense of electromagnetism as we shall
see in the next section. Suppose we add to the Lagrangian a term $\frac{df}{dt}-\dot{\lambda}^{t}\nabla f-\frac{\partial f}{\partial t}=0$
where $f(\lambda,t)$ is to be determined. The extra terms allow us
to ``complete the square'' in the Lagrangian as follows: 
\begin{equation}
\frac{1}{2}\left(\dot{\lambda}-\Phi(f,\lambda)\right)^{t}g\left(\dot{\lambda}-\Phi(f,\lambda)\right)+F=\frac{1}{2}\dot{\lambda}^{t}g\dot{\lambda}-\dot{\lambda}^{t}\left(M+\nabla f\right)+\frac{1}{2}\phi-\frac{\partial f}{\partial t}+\frac{df}{dt}\label{factorisation}
\end{equation}
 Equating terms gives the three equations
\begin{eqnarray}
\Phi & = & g^{-1}(\nabla f+M)\label{baseeq1}\\
\frac{1}{2}\Phi^{t}g\Phi & = & \frac{1}{2}\phi-\frac{\partial f}{\partial t}\label{baseeq2}\\
F & = & \frac{df}{dt}
\end{eqnarray}

Now the momenta $p$ corresponding to $\lambda$ and the Friedlin-Wentzel
Hamiltonian $\mathscr{\mathcal{H}}$ are easily computed to be
\begin{eqnarray}
p & = & g\dot{\lambda}-M\nonumber \\
\mathcal{H}(p,\lambda) & = & \dot{\lambda}^{t}p-\mathcal{L}\\
 & = & \frac{1}{2}\dot{\lambda}^{t}g\dot{\lambda}-\frac{1}{2}\phi\\
 & = & \frac{1}{2}\left(p+M\right)^{t}g^{-1}\left(p+M\right)-\frac{1}{2}\phi\label{Hamilt}
\end{eqnarray}

Substitution of (\ref{baseeq1}) into (\ref{baseeq2}) and comparison
with (\ref{Hamilt}) shows that
\begin{equation}
\mathcal{H}(\nabla f,\lambda)+\frac{\partial f}{\partial t}=0\label{HJeq}
\end{equation}

in otherwords the Hamilton-Jacobi (HJ) equation for this Lagrangian.
The non-negative action $S$, which represents the information loss
of a particular path, may now be written instructively as proportional
to the terms
\[
S\propto\int_{0}^{T}\left(\dot{\lambda}-\Phi(f(\lambda,t),\lambda)\right)^{t}g\left(\dot{\lambda}-\Phi(f(\lambda,t),\lambda)\right)dt+f(\lambda(T),T)-f(\lambda_{0},0)
\]
Such an equation holds for many choices for the gauge function $f$
\uline{providing} they satisfy the HJ equation. One interesting
choice for interpreting this reformulation occurs if we specify it
at the endpoint:
\[
f(\lambda,T)=0
\]

The values of $f$ between $t=0$ and $t=T$ which gives $\Phi$,
may be obtained by integrating the HJ equation \uline{back in time}
from the endpoint to the start point. We denote this particular solution
by $f(\lambda,T,t)$. With such a choice and a specified $\lambda_{0}$,
the last two terms for the action become independent of $\lambda(T)$.
It is clear now since $g$ is non-negative definite, that the action
at $t=T$ will be minimized providing that
\begin{eqnarray}
\dot{\lambda} & = & \Phi(f(\lambda,T,t),\lambda)=g^{-1}(\nabla f(\lambda,T,t)+M)\nonumber \\
\lambda(0) & = & \lambda_{0}\label{varyingOet}
\end{eqnarray}

which obviously uniquely specifies a path between $t=0$ and $t=T$.
Such an action minimizing path is not however a thermodynamical path
as may be seen easily by consideration of the very simple example
presented in the previous section. A straightforward calculation shows
for that case that it is actually a extremal path instead. Similarly
if we take the limit $\Delta t\rightarrow\infty$ then the consistency
distribution is determined by the actions of extremal paths only and
thus clearly the path determined by (\ref{varyingOet}) will provide
the maximum of the consistency distribution for $t=T$. No such deduction
is possible for the general $\Delta t$ case however since the consistency
distribution depends then on \uline{all} paths not just the extremals. 

It is clear however from (\ref{varyingOet}) that the resulting Lagrangian
with this choice of gauge $f$ is \uline{not} of a generalized
OM form since the function $\Phi$ obviously depends on the endpoint
$T$ chosen. 

Suppose now instead we choose $f$ to be time independent i.e. a solution
$f_{s}$ of the stationary HJ equation. 
\[
\mathcal{H}(\nabla f_{s},\lambda)=0
\]

The action can now be written as 
\begin{equation}
S(0,T)\propto\int_{0}^{T}\left(\dot{\lambda}-\Phi(f_{s}(\lambda),\lambda)\right)^{t}g\left(\dot{\lambda}-\Phi(f_{s}(\lambda),\lambda)\right)dt+f_{s}(\lambda(T))-f_{s}(\lambda_{0})\label{canonical}
\end{equation}

The first term here is clearly of generalized OM form while the other
terms depend only on the endpoints of the path. Thus the path integral
here is similar to that discussed by \cite{graham1977path} and others
but here the additional terms beyond the generalized OM action depend
only on path endpoints. In addition $\Phi$ and $g$ may be determined
from first principles not empirically prescribed. The consistency
distribution can now be written as
\begin{equation}
\psi(\lambda_{T})=C\exp\left(-\Delta tf_{s}(\lambda_{T})\right)K_{OM}\left(\lambda_{0},\lambda_{T}\right)\label{OMrelation}
\end{equation}
where $K_{OM}$ is a path integral of generalized OM form. One might
be tempted at this point to eliminate the endpoint function here by
adding to the original action an additional endpoint cost term analogous
to an entropy function as is done effectively in the original work
by OM. A careful consideration however of the simple example of the
previous section shows that this has the effect of eliminating the
spinup plateau effect for the thermodynamical path which is seen to
be essential from DNS studies. 

Further progress in analysis may be made now by considering the large
$\Delta t$ case because $K_{OM}$ has then been considered in depth
by \cite{ventsel1970small}. This limit is commonly referred to as
the weak noise limit. The consistency distribution in that limit is
proportional to a solution of the Fokker Planck equation for a multiplicative
stochastic process with Ito form:
\begin{eqnarray}
d\lambda & = & \Phi dt+\Pi dW\nonumber \\
\Pi^{T}\Pi & = & \left(\Delta tg\right)^{-1}\label{sde}
\end{eqnarray}

This connection is not surprising given the more general work of \cite{graham1977path}
with the same stochastic processes. 

The weak noise limit for such equations have also been extensively
studied in the literature (eg \cite{gard} Chapter 6) using perturbation
expansion methods. To first order in $\epsilon\equiv\frac{1}{\sqrt{\Delta t}}$
the stochastic process becomes time dependent Ornstein Uhlenbeck. More precisely let the solution of 
\begin{eqnarray}
\dot{\lambda} & = & \Phi(\lambda)\label{oet}\\
\lambda(0) & = & \lambda_{0}\nonumber 
\end{eqnarray}

be denoted by $\alpha(t)$ and define the vector variable
\[
y=\frac{\lambda-\alpha}{\epsilon}
\]

which is a rescaled deviation from $\alpha(t)$. To first order in
$\epsilon\equiv\frac{1}{\sqrt{\Delta t}}$ the stochastic process
becomes time dependent Ornstein Uhlenbeck with drift vector
and noise covariance matrix
\begin{eqnarray*}
A_{i}(t,y) & = & \frac{\partial\Phi_{i}}{\partial\lambda_{j}}\left(\lambda=\alpha(t)\right)y_{j}\\
B_{ij}(t) & = & g^{-1}\left(\lambda=\alpha(t)\right)
\end{eqnarray*}
.Let us now consider the limit of large time and assume that in that
case $\alpha(t)\rightarrow\alpha^{*}$. For such large times, deviations
of $\lambda$ from $\alpha^{*}$ form a regular multivariate Ornstein
Uhlenbeck process with linear drift vector and constant noise covariance
matrices of approximately
\begin{eqnarray*}
D_{i}^{*}(y) & = & \frac{\partial\Phi_{i}}{\partial\lambda_{j}}\left(\lambda=\alpha^{*}\right)y_{j}\\
B_{ij}^{*} & = & g^{-1}\left(\lambda=\alpha^{*}\right)
\end{eqnarray*}

The large time behaviour of the consistency distribution in this case
may thus be written using (\ref{OMrelation}) approximately as
\[
\psi(\lambda)\approx C\exp\left(-\frac{1}{2}\left(\lambda-\alpha(t)\right)^{t}\sigma^{-1}\left(\lambda-\alpha(t)\right)-\Delta tf_{s}(\lambda)\right)
\]

where $\sigma$ is the steady-state Ornstein Uhlenbeck covariance matrix
obtainable from drift and noise covariance matrices. The maximum $\hat{\lambda}$
of this then determines the thermodynamical path and can be written
as 
\begin{equation}
\hat{\lambda}_{i}=\alpha_{i}(t)-\sigma\frac{\partial f_{s}}{\partial\lambda_{i}}\left(\hat{\lambda}\right)\label{maximumcond}
\end{equation}

At $\alpha^{*}$ it follows from (\ref{oet}) that we have using the
definition of $\Phi$ that
\[
M_{i}(\alpha^{*})+\frac{\partial f_{s}}{\partial\lambda_{i}}\left(\alpha^{*}\right)=0
\]

which implies since $f_{s}$ is a solution of the stationary HJ equation
that $\phi(\alpha^{*})=0$. However from section 2 it follows from
(\ref{ILrev}) and (\ref{decomp}) that 
\[
\phi\geq M^{t}g^{-1}M\geq0
\]

and hence that $M_{i}(\alpha^{*})=\frac{\partial f_{s}}{\partial\lambda_{i}}\left(\alpha^{*}\right)=0$.
Thus one may write in the vicinity of $\alpha^{*}$ that
\[
\frac{\partial f_{s}}{\partial\lambda_{i}}\left(\hat{\lambda}\right)\approx G_{ij}\hat{\lambda}_{j}
\]

with $G$ symmetric. Solving (\ref{maximumcond}) we obtain for the
thermodynamical path
\[
\hat{\lambda}(t)\approx\left(I+\sigma G\right)^{-1}\alpha(t).
\]

Now $\alpha(t)$ satisfies a linearization of the equation (\ref{oet})
about $\alpha^{*}$ which is of the type considered by {\"O}ttinger (see
\cite{ottinger2005beyond}) and is relevant to a large number of practical
non-equilibrium thermodynamical systems.

\section{Connection to the motion of a charged particle in an external electromagnetic
field}

\label{sec:5}

The original form of the Lagrangian (\ref{TurlLagrangian}) is familiar
from classical mechanics. Indeed if we set $g_{ij}=m\delta_{ij}$
then the Hamiltonian from equation (\ref{Hamiltonian}) is identical
with that of a non-relativistic particle moving in an external fixed
electromagnetic field. Here $-M$ and $-\phi$, which generate the
reversible and irreversible flows, are proportional to the magnetic
vector potential%
\footnote{We use $M$ to denote the magnetic vector potential to avoid confusion
with the resolved variable set $A$%
} and the scalar potential%
\footnote{This can include both an electric potential and other potentials such
as gravitation%
} respectively (see \cite{landau1981quantum} p421). 

The more general case for the Fisher metric tensor $g$ is also interesting.
Here the Euler-Lagrange equations corresponding to the Lagrangian
(\ref{TurlLagrangian}) take after a straightforward calculation the
following forced geodesic form
\begin{equation}
\ddot{\lambda}_{l}+\dot{\lambda}_{k}\dot{\lambda}_{l}\Gamma_{kl}^{l}=g^{li}\left[\dot{\lambda}_{k}\left(\frac{\partial M_{i}}{\partial\lambda_{k}}-\frac{\partial M_{k}}{\partial\lambda_{i}}\right)+\frac{\partial\phi}{\partial\lambda_{i}}\right]\label{ELgeodesic}
\end{equation}

where $\Gamma$ is the Christoffel symbol corresponding to the Riemannian
metric tensor $g$ (and the summation convention is assumed). Such
equations are similar in form to the geodesic equations for a particle
subject to an external electromagnetic field within a general space-time
manifold (see \cite{wald84} pp41 and 69) which read 
\begin{eqnarray}
\frac{du^{a}}{d\tau}+u^{c}u^{d}\widetilde{\Gamma}_{cd}^{a} & = & \frac{q}{m}g^{ab}F_{bc}u^{c}\label{grelectromagnetism}\\
u^{a} & \equiv & \frac{dx^{a}}{d\tau}\nonumber \\
F_{ab} & \equiv & \nabla_{a}M_{b}-\nabla_{b}M_{b}=\partial_{a}M_{b}-\partial_{b}M_{a}.\nonumber 
\end{eqnarray}

The tensor indices here are on space-time; the Christoffel symbol
$\tilde{\Gamma}$ is appropriate for the usual Lorentzian (as opposed
to Riemannian) space-time manifold; $\tau$ is the proper time for
the charged particle and finally the electromagnetic potential 4-vector
$M_{a}$ is the combined 3-vector potential and the scalar potential.
Note that the electromagnetic field $F$ can be defined from the potential
using an arbitrary derivative operator not just the covariant derivative
corresponding to the metric since it is the exterior derivative of
the potential. If we assume that the particle is moving non-relativisitically
then we have in a suitable co-ordinate system that 
\begin{equation}
u^{0}=\frac{dx^{0}}{d\tau}\simeq constant>>u^{i}\qquad i=1,2,3\label{non-relativistic}
\end{equation}
Finally if we assume that the space-time is static then we can choose
an appropriate co-ordinate frame%
\footnote{Set by the static space-time Killing vector%
} in which the metric tensor is Riemannian with respect to the spatial
co-ordinates; the cross terms $g_{0i}$ vanish and further
\[
g_{00}=V(x^{1},x^{2},x^{3})
\]
(see \cite{wald84} p119). With respect to the spatial indices, the
left hand side of the geodesic equations are now the same as our Riemannian
version (\ref{ELgeodesic}) with the exception of terms deriving from
cross spatial-temporal Christoffel symbols $\tilde{\Gamma}_{00}^{i}=\frac{\partial V}{\partial x^{i}}$.
Using the nonrelativistic approximation (\ref{non-relativistic})
it is clear this term can be moved to the right hand side and included
in the gradient of the scalar potential. Finally if we assume that
the external electromagnetic field is static as well as the space-time
then equations (\ref{ELgeodesic}) and (\ref{grelectromagnetism})
are easily seen to be of the same form.

\section{Corresponding Schr{\"o}dinger equations }

\label{sec:6}

There is considerable discussion in the literature as to the exact
relationship between the Onsager-Machlup path integral discussed above
and a corresponding Euclidean Schr{\"o}dinger equation for the transition
probability. The interested reader is also referred to the book \cite{chaichian2001path}
where the connection with the issue of quantum operator ordering in
Hamiltonians is explained. In general, the relationship depends on
the precise nature of the temporal limiting process adopted in defining
the path integral. Differing temporal discretisations%
\footnote{Or Fourier/phase decompositions.%
} of quantities within the Lagrangian lead to Schr{\"o}dinger equations
with different drift and potential terms. This ambiguity could be
seen as somewhat academic since it depends on taking the limit $\Delta t\rightarrow0$
which violates the spirit of working on a slow timescale (further
discussion of such a viewpoint can be found in \cite{lavenda1979validity}
in the context of general stochastic processes). Nevertheless, a unique
fully covariant correspondence has been given in \cite{graham1977path}
(see also \cite{dekker1980path}). Graham and co-workers show how
this can be achieved concretely by an appropriately chosen discretisation
procedure motivated by Wilson's renormalization group (see \cite{deininghaus1979nonlinear}
and \cite{dekker1980path}). We follow the Graham formalism below.

We begin for pedagogical reasons with consideration of the simple
case at the beginning of the last section with $g_{ij}=m\delta_{ij}$
namely a charged particle in a flat space with an externally prescribed
electromagnetic field. The Feynman path integral of this system is
very well known and important (\cite{feynman1965quantum} p79) and
the wave function ($\psi(\lambda)\equiv K(\lambda,\lambda_{fixed})$)
satisfies a Schr{\"o}dinger equation discussed at length in standard texts
such as \cite{landau1981quantum}. Formally the derivation of this
equation from the path integral proceeds identically in our case with
the identification
\[
\Delta t\longleftrightarrow-\frac{i}{\hbar}
\]

and so we obtain the Schr{\"o}dinger%
\footnote{This is strictly a Wick rotated Schr{\"o}dinger equation i.e. a parabolic
PDE of a diffusion-absorbtion type.%
} equation
\[
\frac{1}{\Delta t}\frac{\partial\psi}{\partial t}=\frac{1}{2m}\left(\frac{1}{\Delta t}\nabla-M\right)^{t}\left(\frac{1}{\Delta t}\nabla-M\right)\psi-\frac{1}{2}\phi\psi
\]

Noteworthy here is that this formal derivation assumes that in the
path integral Lagrangian, the electromagnetic fields are evaluated
at the midpoint of the time interval used to define the time derivatives.
If other choices are made then a different equation results (see \cite{chaichian2001path}).
In the quantum case these alternate equations do not exhibit gauge
invariance so are ruled out.

\cite{graham1977path} derived the following path integral Lagrangian
\begin{eqnarray*}
L_{g}(\lambda) & = & \frac{1}{2}Q_{ij}^{-1}\left(\dot{\lambda}_{i}-\omega_{i}\right)\left(\dot{\lambda}_{j}-\omega_{j}\right)+\frac{1}{2}\sqrt{Q}\frac{\partial}{\partial\lambda_{k}}\left(\frac{\omega_{k}}{\sqrt{Q}}\right)-V+\frac{1}{12}R\\
\sqrt{\left|Q\right|} & \equiv & \sqrt{\det(Q)}\\
\omega_{i} & \equiv & K_{i}-\frac{1}{2}\sqrt{\left|Q\right|}\frac{\partial}{\partial\lambda_{k}}\left(\frac{Q_{ik}}{\sqrt{\left|Q\right|}}\right)\\
R & = & Riemann\; scalar\; of\; Q^{-1}
\end{eqnarray*}
from the Schr{\"o}dinger equation
\begin{equation}
\left(\frac{1}{2}\frac{\partial}{\partial\lambda_{i}}\frac{\partial}{\partial\lambda_{j}}Q_{ij}-\frac{\partial}{\partial\lambda_{k}}K_{k}+V\right)\psi=\psi_{t}\label{genschrod}
\end{equation}

Comparison of this Lagrangian with one derived earlier gives the following
identifications
\begin{eqnarray*}
Q & = & g^{-1}\\
K_{i} & = & g_{ij}^{-1}M_{j}+\frac{1}{2\sqrt{\left|g\right|}}\frac{\partial}{\partial\lambda_{k}}\left(\sqrt{\left|g\right|}g_{ik}^{-1}\right)\\
V & = & \frac{1}{2\sqrt{\left|g\right|}}\frac{\partial}{\partial\lambda_{k}}\left(\sqrt{\left|g\right|}g_{kl}^{-1}M_{l}\right)+\frac{1}{12}R-\frac{1}{2}\phi+\frac{1}{2}g_{ij}^{-1}M_{i}M_{j}
\end{eqnarray*}

which when substituted in (\ref{genschrod}) gives the appropriate
equation for the current application. In order to make this precise
identification of a Schr{\"o}dinger equation the prescription of \cite{deininghaus1979nonlinear}
for the limit $\Delta t\rightarrow0$ must be assumed. To reiterate,
these equations really are only approximate asymptotic relations given
that on physical grounds $\Delta t$ must be bounded below by the
fast time scale of the dynamical system under consideration.

\section{Relationship to quantum statistical mechanics.}

\label{sec:7}
The most familiar application of path integrals to statistical mechanics
is that which gives the density matrix for an equilibrium ensemble
of quantum states i.e. describes a mixed quantum state (see \cite{feynman1965quantum}
Chapter 10). The path integral then has a Lagrangian which is the
Wick rotation of the classical Lagrangian with imaginary time associated
with inverse temperature. For the case of a particle moving in an
electromagnetic vector potential and a scalar potential as discussed
in section 5 the effect of the Wick rotation is to reverse the sign
of the scalar potential and make the vector (magnetic) potential term
purely imaginary. Without the magnetic potential the resulting path
integral is Wiener and consequently able to be practically evaluated
(see e.g. \cite{ceperley1995path} for application to Bose condensates).
The present path integral is entirely analogous except the magnetic
term in the Lagrangian (which is associated with reversible trajectories)
is real rather than imaginary. In addition the scalar potential $\phi$
is exactly the same i.e. there is no reversal of sign as in the classical
analog of Section 5. This is important since from equations (\ref{ILnew})
and (\ref{order1}) it is easily deduced that $\phi\geq0$ and the
potential achieves the lower bound when $\lambda=0$ i.e. when the
trial density is a Gibbs density. Such lower bounded potentials are
of course common in many different dynamical contexts. The very simple
example discussed above in Section 3 is obviously the density matrix
for an ensemble of particles in a harmonic potential. A major practical
advantage of the present path integral is clearly that it is always
Wiener and thus likely amenable to the numerical methods widely used
in quantum statistical mechanics when there is no magnetic potential.

\section{Existence of a unique steady-state consistency distribution\label{sec:Existence-of-a}}

\label{sec:8}

Consider a time transfer operator $K$ of consistency distributions
for one timestep $\Delta t$. If we choose to time discretise on the
backward timestep then we may write
\begin{eqnarray}
\psi(t+\Delta t,\lambda) & = & K\psi(t,\lambda)=\int R(\lambda,\kappa)\psi(t,\kappa)d\kappa\nonumber \\
R(\lambda,\kappa) & \equiv & N(\kappa)\exp\left[-\frac{\left(\Delta t\right)^{2}}{2}\left(\frac{\lambda-\kappa}{\Delta t}-g^{-1}M\right)^{t}g\left(\frac{\lambda-\kappa}{\Delta t}-g^{-1}M\right)-IL_{rev}\right]\nonumber \\
\label{transferop}
\end{eqnarray}

where all the functions in the exponent are of $\kappa$ the backward
variable rather than $\lambda$. We also choose the function $N$
as
\[
N(\kappa)=\left[2\pi\right]^{-m/2}\sqrt{\left|g\right|}
\]

where $m$ is the coarse grained dimension. This choice for normalization
has the attractive property that it ``preserves volumes'' in $\kappa$
space i.e. $ $$\sqrt{\left|g\right|}d\kappa$ is the natural volume
element for the metric tensor $g$ (see e.g. Appendix B \cite{wald84}). 

The operator $K$ as chosen above turns out to be compact with the
addition of some sufficiency conditions. An operator is compact if
the image of any bounded set is totally bounded (see \cite{conway1990course}).
The Kolmogorov-Riesz theorem on totally bounded sets of $L_{1}$ spaces
asserts (see \cite{HH10}) that as well as the image of $K$ being
bounded we also require that for every $\epsilon>0$ there exists
a $R(\epsilon)$ such that for all $K\psi$ 
\begin{equation}
\intop_{\left|\lambda\right|>R(\epsilon)}\left|K\psi\right|d\lambda<\epsilon\label{totally}
\end{equation}

and secondly that for every $\epsilon>0$ there exist some $\rho>0$
such that for all $ $$K\psi$ and $\gamma$ with $\left|\gamma\right|<\rho$
\begin{equation}
\int d\lambda\left|K\psi(\lambda+\gamma)-K\psi(\lambda)\right|<\epsilon\label{label}
\end{equation}

We have the following theorem the proof of which is rather technical
and may be found in Appendix B:\medskip{}

\noindent \textbf{\large{}Theorem: }If the transfer operator $K$
defined by (\ref{transferop}) satisfies the conditions
\begin{enumerate}
\item $IL_{rev}(\kappa)\rightarrow\infty$ as $\left|\kappa\right|\rightarrow\infty$. 
\item In any bounded region $\left|\kappa\right|\leq M$ $\left|g\right|$
is bounded below and $g^{-1}M$ is bounded above by the usual $R^{m}$
norm. 
\end{enumerate}
then it is compact.

Condition 1. here is the most significant. Such a property holds for
the practical cases examined to date by the author. It corresponds
with the quantum case of an infinite confining potential which is
widely relevant. Note that this potential is not the scalar potential
$\phi$ rather it is $IL_{rev}$ given by equation (\ref{ILrev}).
As was observed in section 2 the absence of this term means the reversible
trajectory is an extremal and the Lagrangian reduces to Onsager-Machlup
form. Thus in some sense this term is fundamentally responsible for
irreversibility.

Consider now the cone $\boldsymbol{C}$ of non-negative functions
belonging to the $L_{1}$ Banach space of real functions. Suppose
$\psi\in\boldsymbol{C}$ and that 
\[
\int R(\lambda_{0},\kappa)\psi(\kappa)d\kappa=0
\]

for fixed $\lambda_{0}$. It follows that $\left\Vert Q\psi(\lambda_{0})\right\Vert _{1}=0$
which implies (see \cite{lieb2001analysis} Chapter 2) that $R(\lambda_{0},\kappa)\psi(\kappa)$
considered as a function of $\kappa$ vanishes almost everywhere in
the Lebesgue measure. But since $R$ is strictly positive everywhere
this must imply that $\psi$ also vanishes almost everywhere in the
Lebesgue measure i.e. it is part of the zero equivalence class of
$L_{1}$ functions. Thus the only functions belonging to $\boldsymbol{C}$
mapped by the operator $K$ to the boundary of the cone are those
that are zero in the sense of the $L_{1}$ space. Re-expressed: The
compact operator $K$ is strongly positive in that all members of
$\boldsymbol{C}$ apart from the zero function class are mapped by
$K$ into its interior. 

Thus all conditions for the Krein-Rutman Theorem (a generalization
of the better known Perron-Frobenius theorem to Banach spaces) are
met (see Theorem 1.2 \cite{du2006order}) which implies that $K$
has a unique%
\footnote{Up to a scalar multiple and the addition of a function vanishing almost
everywhere with respect to the Lebesgue measure. %
} eigenvector belonging to $\boldsymbol{C}$ with a positive eigenvalue.
Any other eigenvalue cannot be positive and must have an eigenvector
outside $\boldsymbol{C}$. This unique eigenvector can clearly be
identified with a unique steady-state consistency distribution.

\section{Discussion and future work}

\label{sec:9}
In the present work we have argued that a macrostate is best described
by a time evolving consistency distribution over a trial density manifold.
The distribution may be written as a path integral over the set of
all paths leading to the final manifold location. The maximum of this
consistency distribution defines the best approximating trial density
for the non-equilibrium system and is referred to as the thermodynamical
path. This also specifies via a Legendre transformation, the approximate
expectation values of the slow variables of the system which are the
practical quantites of interest. The complete consistency distribution
is required to describe the dynamical evolution of macrostates but
only the maximum is required for an identification of the best approximate
expectation values. In this (and other) respects the situation is
analogous to quantum mechanics where the complex modulus is observable
statistically but the complex phase is required as well to specify
the evolution of the physical quantum state. It remains a topic for
further research as to whether more of the consistency distribution
beyond the maximum could be used to deduce further information of
practical interest such as a measure of the uncertainty of the slow
variable expectation values derived.

We saw in section 4 that the path integral is one of modified generalized
Onsager-Machlup form. For large values of the slow timescale $\Delta t$
the thermodynamical path can be shown to be related to one of {\"O}ttinger
form (see \cite{ottinger2005beyond}) and indeed as $t\rightarrow\infty$
is directly of such a form up to a matrix multiplication. BT also
discussed a non-stationary formulation for thermodynamical paths for
this situation which \uline{directly} satisfied an {\"O}ttinger equation
of a different type. There the irreversible part of the {\"O}ttinger equation
was time dependent in constrast to the present situation where it
is fixed but the consistency distribution has an additional endpoint
factor. It would be very interesting to directly compare the two thermodynamical
paths since the BT theory is appropriate for large $\Delta t$ and
has worked well in various DNS cases.

A key practical advantage of the present formulation lies in the fact
that extensively tested numerical methods from equilibrium quantum
statistical mechanics exist for the efficient numerical evaluation
of the proposed path integrals. A highly detailed review of this field
from the viewpoint of quantum chemistry may be found in the article
by Ceperley \cite{ceperley1995path}. The techniques therein are currently
being applied by the author to investigate the accuracy of the present
formalism in a series of realistic statistical systems. 

Another issue requiring further investigation concerns the choice
of resolved variables. These are functions of the slow variables of
the original dynamical system but the key question is their specific
selection. Intuitively one expects the densities for random variables
averaged over the time interval $\Delta t$ to be rather general functions
of the slow variables of the original system. Practical experience
however shows that only rather simple such functions are needed when
direct numerical simulations are examined. Thus, for example, the
author has examined the TBH system discussed in section 1 and discovered
that to a very good approximation the square of slow variables suffices
in addition to linear functions. Clearly then an important topic to
examine is the convergence of results from the present formalism as
higher order slow variable functions are included among the slow variables.
Conceptually this can be viewed as refining the trial density manifold
and examining the consequent convergence of the expectation values
of important functions of the slow variables.

The slow variable averaging interval $\Delta t$ used in the proposed
generalized Boltzmann principle also deserves further investigation
in the same way. It would be interesting to document the sensitivity
of slow variable expectations to variations in this parameter. It
seems clear however that the value of this parameter should be set
physically at least approximately by the maximum time scale required
for fast variables to decorrelate.
\section*{Appendix A: Some useful relations}

Define the expectation bracket
\[
\left\langle F\right\rangle \equiv\int F\hat{p}
\]

for a general function of the state variables and time $F$. We have
now
\begin{eqnarray}
\frac{\partial\left\langle F\right\rangle }{\partial t}-\left\langle L(F)\right\rangle  & = & \left\langle \frac{\partial F}{\partial t}\right\rangle +\int(F\hat{p}_{t}-L(F)\hat{p})\nonumber \\
 & = & \left\langle \frac{\partial F}{\partial t}\right\rangle +\int(F(\partial_{t}+L)\hat{p}\nonumber \\
 & = & \left\langle \frac{\partial F}{\partial t}\right\rangle +\int FR\hat{p}\nonumber \\
 & = & \left\langle \frac{\partial F}{\partial t}+FR\right\rangle \label{evol}
\end{eqnarray}

where we are using the anti-Hermitian nature of $L$ on the second
line. Setting $F=1$ we obtain immediately that 
\begin{equation}
\left\langle R\right\rangle =0\label{r0}
\end{equation}

For an exponential family $\hat{p}$ it follows from the definition
(\ref{residual}) that

Now it is easily derived from the definition of $R$ and the form
of the exponential family of distributions that 
\begin{equation}
R=\dot{\lambda}^{t}U+\lambda{}^{t}LA\label{resida}
\end{equation}
which when combined with (\ref{r0}) yields
\begin{equation}
\lambda^{t}\left\langle LA\right\rangle =0\label{order1}
\end{equation}

The anti-Hermitian nature of $L$ also allows us to deduce the following
two useful relations (using the summation convention and vector/matrix
indices for clarity):
\begin{eqnarray}
M_{i}=\left\langle L(A_{i})\right\rangle  & = & \int L(A_{i})\exp\left(\lambda_{j}A_{j}-G\right)\exp\left(-\beta E\right)\nonumber \\
 & = & \int L\left(A_{i}\exp\left(\lambda_{j}A_{j}-G\right)\right)\exp\left(-\beta E\right)-\int A_{i}L\left(\exp\left(\lambda_{j}A_{j}-G\right)\right)\exp\left(-\beta E\right)\nonumber \\
 & = & -\int A_{i}\exp\left(\lambda_{j}A_{j}-G\right)L\left(\exp\left(-\beta E\right)\right)-\lambda_{j}\int A_{i}L\left(A_{j}\right)\hat{p}\nonumber \\
 & = & -\lambda_{j}\left\langle A_{i}L\left(A_{j}\right)\right\rangle =-\lambda_{j}\left\langle (A_{i}-a_{i})L\left(A_{j}\right)\right\rangle \equiv-h_{ij}\lambda_{j}\label{heq}
\end{eqnarray}

where we are using the fact that $L$ annihilates $E$ and (\ref{order1})
for the second last step. Combining (\ref{order1}) and (\ref{heq})
we obtain
\[
\lambda^{*}h\lambda=0
\]
 In a completely analogous way to (\ref{heq}) we deduce that
\[
\left\langle L^{2}(A_{i})\right\rangle =-\lambda_{j}\left\langle LA_{j}LA_{i}\right\rangle \equiv-k_{ij}\lambda_{j}.
\]

and more generally
\begin{equation}
\left\langle L^{n}A_{j}\right\rangle =-\lambda_{i}\left\langle LA_{i}L^{n-1}A_{j}\right\rangle \label{Ln}
\end{equation}

It is easily shown also that 

\[
\frac{\partial M_{i}}{\partial\lambda_{j}}=\left\langle L(A_{i})\left(A_{j}-a_{j}\right)\right\rangle =h_{ji}
\]

\section*{Appendix B: Section \ref{sec:Existence-of-a} Theorem proof.}

We first establish that the operator $K$ is bounded with respect
to the $L_{1}$ norm. Consider the effect of $K$ on a distribution
$\phi$ with $L_{1}$ norm unity:
\begin{eqnarray*}
OP\equiv\left\Vert K\psi\right\Vert _{1} & = & \int\left|\int R\psi d\kappa\right|d\lambda\\
 & \leq & \int\left[\int\left|R\psi\right|d\kappa\right]d\lambda\\
 & = & \iint N\exp\left[-IL_{irr}-IL_{rev}\right]\left|\psi\right|d\kappa d\lambda\\
 & \leq & \iint N\exp\left[-IL_{irr}\right]\left|\psi\right|d\kappa d\lambda\\
 & = & \int\left|\psi\right|d\kappa=1
\end{eqnarray*}

where on line 4 we have used the fundamental fact derived in section
2 that $IL_{rev}\geq0$ while the last line follows after switching
variables of integration and using the normalization condition which
also holds for $\exp\left[-IL_{irr}\right]$$ $.

We further establish that the image of $K$ is totally bounded which
means establishing the additional two properties (\ref{totally})
and (\ref{label}). 

From condition 1. of the Theroem we deduce that there exists $ $a
$\left|\kappa_{0}\right|$ such that 
\[
\exp\left(-IL_{rev}\right)<\frac{\epsilon}{2}\qquad if\;\left|\kappa\right|>\left|\kappa_{0}\right|
\]

Consider now the bounded region $\left|\kappa\right|\leq\left|\kappa_{0}\right|$.
From condition 2. of the Theorem; the region boundedness and the fact
that $\exp\left(-IL_{irr}\right)$ is Gaussian in $\lambda$, if follows
that there exists an $R(\epsilon)>0$ such that for all $\left|\kappa\right|\leq\left|\kappa_{0}\right|$
\[
\intop_{\left|\lambda\right|>R(\epsilon)}N\exp\left(-IL_{irr}\right)d\lambda<\frac{\epsilon}{2}
\]

Thus
\begin{eqnarray*}
\intop_{\left|\lambda\right|>R(\epsilon)}\left|K\psi\right|d\lambda & = & \intop_{\left|\lambda\right|>R(\epsilon)}d\lambda\left|\int d\kappa N\exp\left(-IL_{irr}-IL_{rev}\right)\psi\right|\\
 & \leq & \intop_{\left|\lambda\right|>R(\epsilon)}d\lambda\int d\kappa N\exp\left(-IL_{irr}-IL_{rev}\right)\left|\psi\right|\\
 & \leq & \intop_{\left|\lambda\right|>R(\epsilon)}d\lambda\intop_{\left|\kappa\right|\leq\left|\kappa_{0}\right|}d\kappa N\exp\left(-IL_{irr}\right)\left|\psi\right|\\
 &  & \qquad\qquad+\intop_{\left|\kappa\right|>\left|\kappa_{0}\right|}d\kappa\exp\left(-IL_{rev}\right)\left|\psi\right|\\
 & < & \frac{\epsilon}{2}+\frac{\epsilon}{2}
\end{eqnarray*}

which establishes (\ref{totally}). 

To establish the other required property consider an arbitrary $\rho_{t}>0$
and all $\gamma$ with $\left|\gamma\right|<\rho_{t}$. The triangle
inequality plus (\ref{totally}) implies that 
\begin{eqnarray}
\intop_{\left|\lambda\right|>R(\frac{\epsilon}{4})+\rho_{t}}\left|K\psi\left(\lambda+\gamma\right)-K\psi(\lambda)\right|d\lambda & \leq & \intop_{\left|\lambda\right|>R(\frac{\epsilon}{4})+\rho_{t}}\left|K\psi\left(\lambda+\gamma\right)\right|d\lambda+\intop_{\left|\lambda\right|>R(\frac{\epsilon}{4})+\rho_{t}}\left|K\psi(\lambda)\right|d\lambda\nonumber \\
 & \leq & \frac{\epsilon}{4}+\frac{\epsilon}{4}\label{ineq2}
\end{eqnarray}

Set $S(\epsilon,\rho)=R(\frac{\epsilon}{4})+\rho$ and $V_{\epsilon\rho_{t}}$
the volume of the region $Z:\left|\lambda\right|\leq S(\epsilon,\rho_{t})$.
Let $\kappa_{0}$ be such that $\left|\kappa\right|>\left|\kappa_{0}\right|$
implies that 
\begin{equation}
\exp\left(-IL_{rev}\right)<\frac{\epsilon}{8V_{\epsilon\rho_{t}}}\label{ineq3}
\end{equation}
We have

\begin{eqnarray}
\intop_{\left|\lambda\right|\leq S(\epsilon.\rho_{t})}\left|K\psi\left(\lambda+\gamma\right)-K\psi(\lambda)\right|d\lambda\nonumber \\
=\intop_{\left|\lambda\right|\leq S(\epsilon.\rho_{t})}\intop_{\left|\kappa\right|>\left|\kappa_{0}\right|}\exp\left(-IL_{red}\right) & N\left|\exp\left(-IL_{irr}(\lambda+\gamma\right)-\exp\left(-IL_{irr}(\lambda\right)\right|\left|\psi\right|d\kappa d\lambda\nonumber \\
+\intop_{\left|\lambda\right|\leq S(\epsilon.\rho_{t})}\left|f_{\kappa_{0}}(\lambda+\gamma)-f_{\kappa_{0}}(\lambda)\right|d\lambda\label{ineq3-1}
\end{eqnarray}

with
\[
f_{\kappa_{0}}\left(\lambda\right)\equiv\intop_{\left|\kappa\right|\leq\left|\kappa_{0}\right|}R(\lambda,\kappa)\psi(\lambda)d\kappa
\]

an integral transform defined on a bounded domain. The first integral
on the RHS of (\ref{ineq3-1}) is easily shown using the triangle
inequality; the inequality (\ref{ineq3}) and the non-negativity of
the $IL$ terms to be less than $\frac{\epsilon}{4}$.

The function $f_{\kappa_{0}}$ can be shown by standard arguments
to be continuous since the integral transform is defined on a bounded
domain and the function $R$ is continuous with respect to the first
argument. By the Heine-Cantor theorem it is therefore uniformly continuous
on the bounded region $Z$. It follows that there exists a $\rho_{U}$
such that for all $\gamma:\left|\gamma\right|\leq\rho_{U}$ and all
$\lambda\in Z$
\begin{equation}
\left|f_{\kappa_{0}}(\lambda+\gamma)-f_{\kappa_{0}}(\lambda)\right|<\frac{\epsilon}{4V_{\epsilon\rho_{t}}}\label{ineq4}
\end{equation}

and so for such $\gamma$ the second integral from (\ref{ineq3-1})
is also less than $\frac{\epsilon}{4}.$ Compare now $\rho_{U}$ and
$\rho_{t}$. If $\rho_{U}\geq\rho_{t}$ then we can replace $\rho_{U}$
with $\rho_{t}$ in the last integral inequality discussed and obtain
the required inequality (\ref{label}) by combining the three inequalities
derived from (\ref{ineq2}) and (\ref{ineq3-1}). Conversely if $\rho_{U}<\rho_{t}$
then $V_{\epsilon\rho}<V_{\epsilon\rho_{t}}$ Thus inequalities (\ref{ineq3})
and (\ref{ineq4}) still hold if we use $\rho_{U}$ in place of $\rho_{t}$.
Furthermore the newly defined bounded $Z$ is a subset of the old
$Z$ whence the uniform continuity just discussed holds with the same
$\rho_{U}$ and hence we are done.

\begin{acknowledgements}
Comprehensive discussions with Bruce Turkington on matters related
to the present contribution are very gratefully acknowledged. Useful
discussions on related matters over many years with Andy Majda are
also acknowledged. This paper is dedicated to my mother Annette.
\end{acknowledgements}


\begin{thebibliography}{10}
\providecommand{\url}[1]{{#1}}
\providecommand{\urlprefix}{URL }
\expandafter\ifx\csname urlstyle\endcsname\relax
  \providecommand{\doi}[1]{DOI~\discretionary{}{}{}#1}\else
  \providecommand{\doi}{DOI~\discretionary{}{}{}\begingroup
  \urlstyle{rm}\Url}\fi

\bibitem{ama00}
Amari, S., Nagaoka, H.: Methods of Information Geometry.
\newblock Translations of Mathematical Monographs, AMS, Oxford University Press
  (2000)

\bibitem{battezzati2012onsager}
Battezzati, M.: Onsager principle for nonlinear mechanical systems modeled by
  stochastic dissipative equations.
\newblock Arch. Mech. \textbf{64}(2), 177--206 (2012)

\bibitem{ceperley1995path}
Ceperley, D.M.: Path integrals in the theory of condensed helium.
\newblock Rev Mod Phys \textbf{67}(2), 279 (1995)

\bibitem{chaichian2001path}
Chaichian, M., Demichev, A.: Path integrals in physics. {V}ol. 1: {S}tochastic
  processes and quantum mechanics.
\newblock IOP, London (2001)

\bibitem{conway1990course}
Conway, J.B.: A course in functional analysis.
\newblock Springer (1990)

\bibitem{cov91}
Cover, T., Thomas, J.: Elements of information theory, 2nd edn.
\newblock Wiley-Interscience, New York (2006)

\bibitem{darve2009computing}
Darve, E., Solomon, J., Kia, A.: Computing generalized {L}angevin equations and
  generalized {F}okker--{P}lanck equations.
\newblock Proc. Nat. Acad. Sci. \textbf{106}(27), 10,884--10,889 (2009)

\bibitem{deininghaus1979nonlinear}
Deininghaus, U., Graham, R.: Nonlinear point transformations and covariant
  interpretation of path integrals.
\newblock Z. Phys. B Con. Mat. \textbf{34}(2), 211--219 (1979)

\bibitem{dekker1980path}
Dekker, H.: On the path integral for diffusion in curved spaces.
\newblock Physica A \textbf{103}(3), 586--596 (1980)

\bibitem{du2006order}
Du, Y.: Order structure and topological methods in nonlinear partial
  differential equations: {V}ol. 1: {M}aximum principles and applications.
\newblock World Scientific Publishing Company (2006)

\bibitem{Ein10}
Einstein, A.: Theorie der {O}paleszenz von homogenen {F}luessigkeiten und
  {F}luessigkeitsgemischen in der {N}aehe des kritischen {Z}ustandes.
\newblock Ann. Physik, \textbf{33}, 1275, (1910)

\bibitem{eyink1996action}
Eyink, G. L.: Action principle in nonequilibrium statistical dynamics.
\newblock Phys. Rev. E, \textbf{54}(4), 3419--3435, (1996)

\bibitem{feynman1965quantum}
Feynman, R.P., Hibbs, A.R., Styer, D.F.: Quantum mechanics and path integrals.
\newblock McGraw-Hill New York (1965)

\bibitem{gard}
Gardiner, C. W.: Handbook of Stochastic Methods for Physics, Chemistry and the
  Natural Sciences.
\newblock Springer, (2004)

\bibitem{graham1977path}
Graham, R.: Path integral formulation of general diffusion processes.
\newblock Z. Phys. B Con. Mat. \textbf{26}(3), 281--290 (1977)

\bibitem{haken1976generalized}
Haken, H.: Generalized {O}nsager-{M}achlup function and classes of path
  integral solutions of the {F}okker-{P}lanck equation and the {M}aster
  equation.
\newblock Z. Phys. B Con. Mat. \textbf{24}(3), 321--326 (1976)

\bibitem{HH10}
Hanche-Olsen, H., Holden, H.: The {K}olmogorov-{R}iesz {C}ompactness {T}heorem.
\newblock Expositiones Mathematicae \textbf{28}, 385--394 (2010)

\bibitem{kleeman2012nonequilibrium}
Kleeman, R., Turkington, B.E.: A nonequilibrium statistical model of spectrally
  truncated {B}urgers-{H}opf dynamics.
\newblock Comm. Pure Appl. Math.  (2013).
\newblock In press

\bibitem{kraichnan1979variational}
Kraichnan, R. H.: Variational method in turbulence theory.
\newblock Phys. Rev. Lett., \textbf{42}(19), 1263--1266, (1979)

\bibitem{landau1981quantum}
Landau, L.D., Lifshitz, E.M.: Quantum mechanics non-relativistic theory.
\newblock Pergamon, London (1965)

\bibitem{lavenda1979validity}
Lavenda, B.H.: On the validity of the {O}nsager-{M}achlup postulate for
  nonlinear stochastic processes.
\newblock Found. Phys. \textbf{9}(5-6), 405--420 (1979)

\bibitem{lieb2001analysis}
Lieb, E.H., Loss, M.: Analysis.
\newblock American Mathematical Society, Providence R.I. (2001)

\bibitem{onsager1953fluctuations}
Onsager, L., Machlup, S.: Fluctuations and irreversible processes.
\newblock Phys. Rev. \textbf{91}(6), 1505--1512 (1953)

\bibitem{ottinger2005beyond}
{\"O}ttinger, H.C.: Beyond equilibrium thermodynamics.
\newblock Wiley-Interscience (2005)

\bibitem{roncadelli1992new}
Roncadelli, M.: New path integral representation of the quantum mechanical
  propagator.
\newblock J. Phys. A-Math. Gen. \textbf{25}(16), L997 (1992)

\bibitem{taniguchi2007onsager}
Taniguchi, T., Cohen, E.G.D.: Onsager-{M}achlup theory for nonequilibrium
  steady states and fluctuation theorems.
\newblock J. Stat. Phys. \textbf{126}(1), 1--41 (2007)

\bibitem{turkington2012optimization}
Turkington, B.: An optimization principle for deriving nonequilibrium
  statistical models of hamiltonian dynamics.
\newblock J. Stat. Phys \textbf{152}, 569--597 (2013)

\bibitem{ventsel1970small}
Ventsel, A.D., Freidlin, M.I.: On small random perturbations of dynamical systems.
\newblock Russ. Math. Surv+, \textbf{25}(1), 1--55, (1970)

\bibitem{wald84}
Wald, R.M.: General Relativity.
\newblock University of Chicago Press, Chicago and London (1984)

\bibitem{Zub74}
Zubarev, D.N.: Nonequilibrium {S}tatistical {T}hermodynamics.
\newblock Plenum Press, New York (1974)

\bibitem{zwanzig2001nonequilibrium}
Zwanzig, R.: {Nonequilibrium statistical mechanics}.
\newblock Oxford University Press, USA (2001)

\end{thebibliography}


\end{document}